\documentclass[a4paper,fleqn,usenatbib]{mnras}

\usepackage{newtxtext,newtxmath}

\usepackage[T1]{fontenc}
\usepackage{ae,aecompl}

\usepackage{graphicx}	
\usepackage{amsmath}	
\usepackage{amssymb}	
\usepackage{braket}
\usepackage{cprotect}

\def\mr#1{\mathrm{#1}}

\title[Sloshing cold front in the Perseus cluster]{Substructures associated with the sloshing cold front in the Perseus cluster}

\author[Y. Ichinohe et al.]{
Y. Ichinohe,$^{1}$\thanks{E-mail: ichinohe@rikkyo.ac.jp}
A. Simionescu,$^{2,3,4}$
N. Werner,$^{5,6,7}$
A. C. Fabian$^{8}$
and T. Takahashi$^{4}$
\\
$^{1}$Department of Physics, Rikkyo University, 3-34-1 Nishi-Ikebukuro, Toshima, Tokyo 171-8501, Japan\\
$^{2}$SRON Netherlands Institute for Space Research, Sorbonnelaan 2, 3584 CA Utrecht, The Netherlands\\
$^{3}$Institute of Space and Astronautical Science, Japan Aerospace Exploration Agency, 3-1-1 Yoshinodai, Chuo, Sagamihara, Kanagawa 252-5210, Japan\\
$^{4}$Kavli IPMU, UTIAS, The University of Tokyo, 5-1-5 Kashiwa-no-Ha, Kashiwa City, Chiba 277-8583, Japan\\
$^{5}$MTA-E\"{o}tv\"{o}s University Lend\"{u}let Hot Universe Research Group, P\'{a}zm\'{a}ny P\'{e}ter s\'{e}t\'{a}ny 1/A, Budapest, 1117, Hungary\\
$^{6}$Department of Theoretical Physics and Astrophysics, Faculty of Science, Masaryk University, Kotlarsk\'{a} 2, Brno, 611 37, Czech Republic\\
$^{7}$School of Science, Hiroshima University, 1-3-1 Kagamiyama, Higashi-Hiroshima 739-8526, Japan\\
$^{8}$Institute of Astronomy, Madingley Road, Cambridge CB3 0HA\\
}

\date{\today}

\pubyear{2018}

\begin{document}
\label{firstpage}
\pagerange{\pageref{firstpage}--\pageref{lastpage}}
\maketitle

\begin{abstract}
 X-ray substructures in clusters of galaxies provide indirect clues about the microphysical properties of the intracluster medium (ICM), which are still not very well known. In order to investigate X-ray substructures in detail, we studied archival $\sim$1~Msec Chandra data of the core of the Perseus cluster, focusing on the substructures associated with the sloshing cold front. In the east half of the cold front, we found a Kelvin-Helmholtz instability (KHI) layer candidate. The measured width-to-azimuthal extension ratio and the thermodynamic properties are all consistent with it being a KHI layer currently developing along the sloshing cold front. We found a thermal pressure deficit of the order of 10$^{-2}~\mr{keV~cm^{-3}}$ at the KHI layer. Assuming that turbulent pressure fully supports the pressure deficit, we estimated the turbulent strength at several hundred km~s$^{-1}$, which translates into the turbulent heating rate of $Q_\mr{turb}\sim 10^{-26}~\mr{erg~cm^{-3}~s^{-1}}$. This value agrees within an order of magnitude with the previous estimation derived from the surface brightness fluctuations, and can balance the radiative cooling at this radius. In the west half of the cold front, we found feather-like structures which are similar to the structures observed in recent numerical simulations of the gas sloshing of magnetized plasma. Their thermodynamic properties are consistent with one of the feathers being a projected gas depletion layer induced by the amplified magnetic field whose strength is $B\sim$30$~\mr{\mu G}$.
\end{abstract}




\section{Introduction}
A significant fraction of the baryons in the present-day Universe is in the form of the intracluster medium (ICM). It shines brightly in X-rays because of its high temperature resulting from the deep gravitational potential of galaxy clusters and merger shocks during their hierarchical growth. One of the most important astrophysical questions regards the microphysical properties of the ICM, which are still not very well known. The complex morphology of X-ray substructures provides indirect clues about the ICM microphysics, because when a dynamical or thermodynamic disturbance, such as a merger or an outburst of the central AGN, occurs in a system, the ICM therein must respond to the disturbance according to its microphysical properties \citep{zuhone16}.

The Perseus cluster is nearby \citep[$z=0.017284$,][]{hitomiv}, massive \citep[$M_{200}=6.65\times10^{14}M_\odot$,][]{simionescu11}, and the brightest cluster of galaxies in the X-ray sky \citep{edge90}. Because of its proximity and X-ray brightness, it has been the most extensively studied galaxy cluster at X-ray wavelengths. The brightness and the depth of observation make the Perseus cluster the best system to study the thermodynamics (i.e. spectroscopic properties) of each substructure to investigate the microphysical properties of ICM.

Many X-ray cavities have been observed around the core of the Perseus cluster. The innermost cavities are filled with radio lobes \citep{boehringer93,churazov00,fabian00}, suggesting that they are bubbles inflated by the jet from the central AGN, being filled with relativistic particles. The outer cavities are ``ghost'' cavities, which are not associated with radio emission peaks \citep{fabian06,fabian11b}. These ghost cavities should still have relativistic plasma, but it has aged and probably only appears at much lower radio frequencies. They are likely related to the past activities of the central AGN.

A deep {\it Chandra} observation of the core of the Perseus cluster revealed the existence of weak shocks and ripples \citep{fabian06}. The ripples seem to be propagating outward, and are thus likely to be sound waves associated with the bubbles, transporting the energy input from the bubble to the ICM \citep{sanders07}.

The X-ray morphology within $\sim$100~kpc from the core is observed to be asymmetric, with a spiral-like arm extending anticlockwise from the vicinity of the centre \citep{churazov00,churazov03,sanders07,fabian11b}. The thermodynamic structures \citep{fabian06} indicate that the spiral is a sloshing cold front \citep{ascasibar06,markevitch07,ichinohe15,ueda17} due to a previous merger. At larger radii, there have also been observations of the substructures associated with the outer sloshing cold front \citep{walker17,walker18}. {\it XMM-Newton} and {\it ROSAT} observations found that the X-ray morphology is asymmetric also on much larger scales \citep{churazov03,simionescu12}, which is probably related to the innermost spiral pattern.

All these surface brightness features represent disturbances that are expected to be seen also in the dynamical properties of the gas. The Perseus cluster is the first-light target that {\it Hitomi} \citep{takahashi16} observed before its communication-loss, and thus is the only galaxy cluster for which an X-ray microcalorimeter \citep[SXS; Soft X-ray Spectrometer;][]{kelley16} observation has been performed \citep{hitomi16}. Using all available Perseus SXS datasets, \citet{hitomiv} mapped the velocity structure around the core out to $\sim$100~kpc. They found that the line-of-sight velocity dispersion reaches maxima of $\sim$200~km~s$^{-1}$ toward the central AGN and toward the northwestern ``ghost'' bubble, and that at the same time the velocity dispersion appears constant around 100~km~s$^{-1}$ everywhere else.

Although a number of observations have been done and many features are investigated in detail from kpc scales to Mpc scales as mentioned above, there are a lot of features which still remain to be studied. Among such structures, we selected two which are seen in the {\it Chandra} image relatively close to the core (within $<$100~kpc from the centre), and both seem to be related to the above-mentioned sloshing cold front, and studied them in detail.

Unless otherwise noted, the error bars correspond to 68\% confidence level for one parameter. Throughout this paper, we assume the standard $\Lambda$CDM cosmological model with the parameters of $(\Omega_m,\Omega_\Lambda,H_0)=(0.3,0.7,70~\mr{km~s^{-1}~Mpc^{-1}})$. In this cosmology, the angular size of 1~arcmin corresponds to the physical scale of 21 kpc at the redshift of $z=0.017284$.

\section{Observations, data reduction, and data analysis}\label{sec:data}
We selected thirteen ObsIDs (3209, 4289, 4946, 4947, 4948, 4949, 4950, 4951, 4952, 4953, 6139, 6145 and 6146; primary focal plane detector: ACIS-S) and reprocessed the archival level 1 event lists produced by the {\it Chandra} pipeline in the standard manner\footnote{CIAO Homepage, Data Preparation; http://cxc.harvard.edu/ciao/threads/data.html} using the {\small CIAO} software package (version 4.10) and the {\small CALDB} version 4.7.8. The resulting total net exposure time is $\sim$1~Msec.

\begin{figure*}
 \begin{minipage}{0.495\hsize}
  \centering
  \includegraphics[width=3.0in]{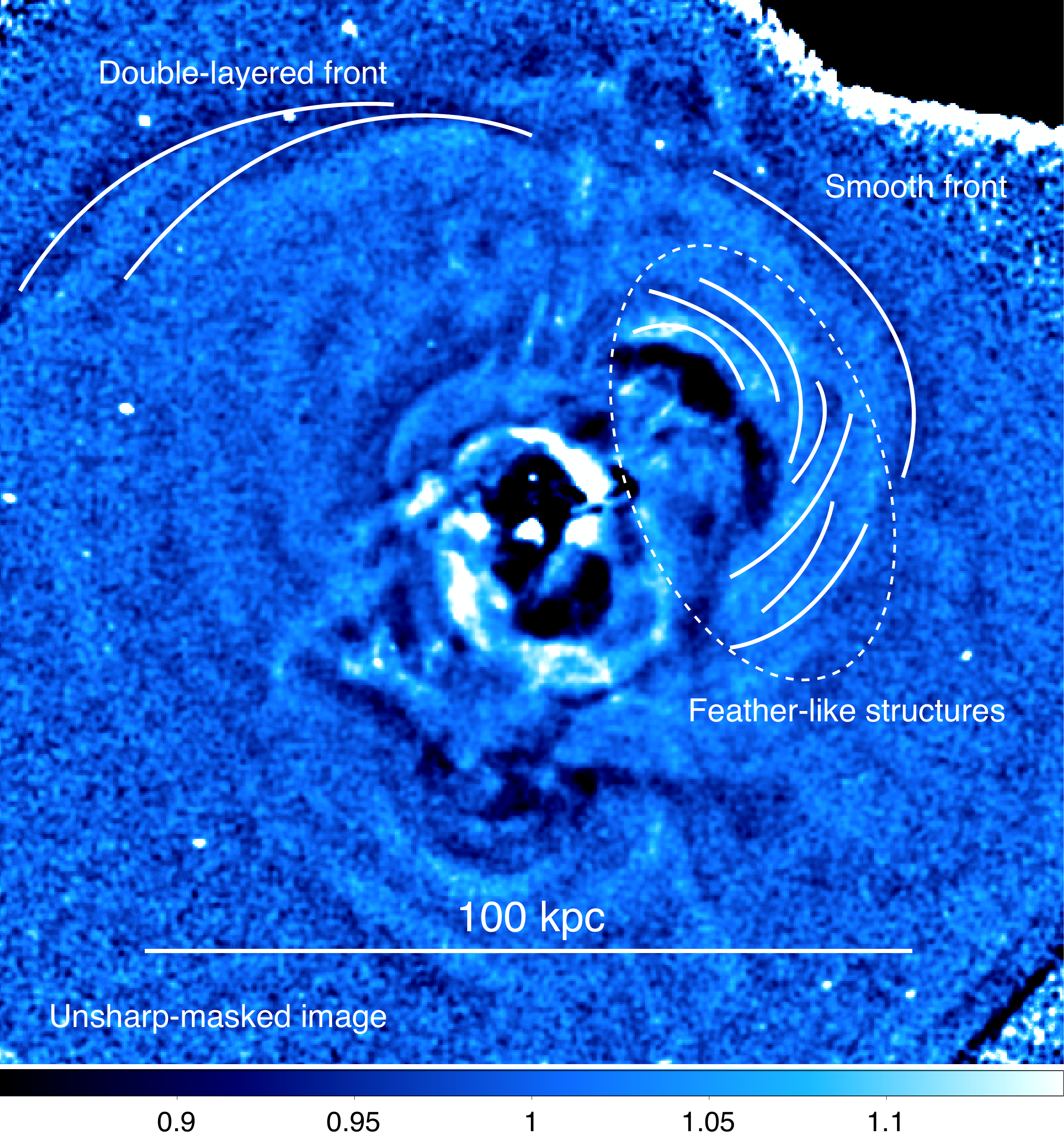}
 \end{minipage}
 \begin{minipage}{0.495\hsize}
  \centering
  \includegraphics[width=3.0in]{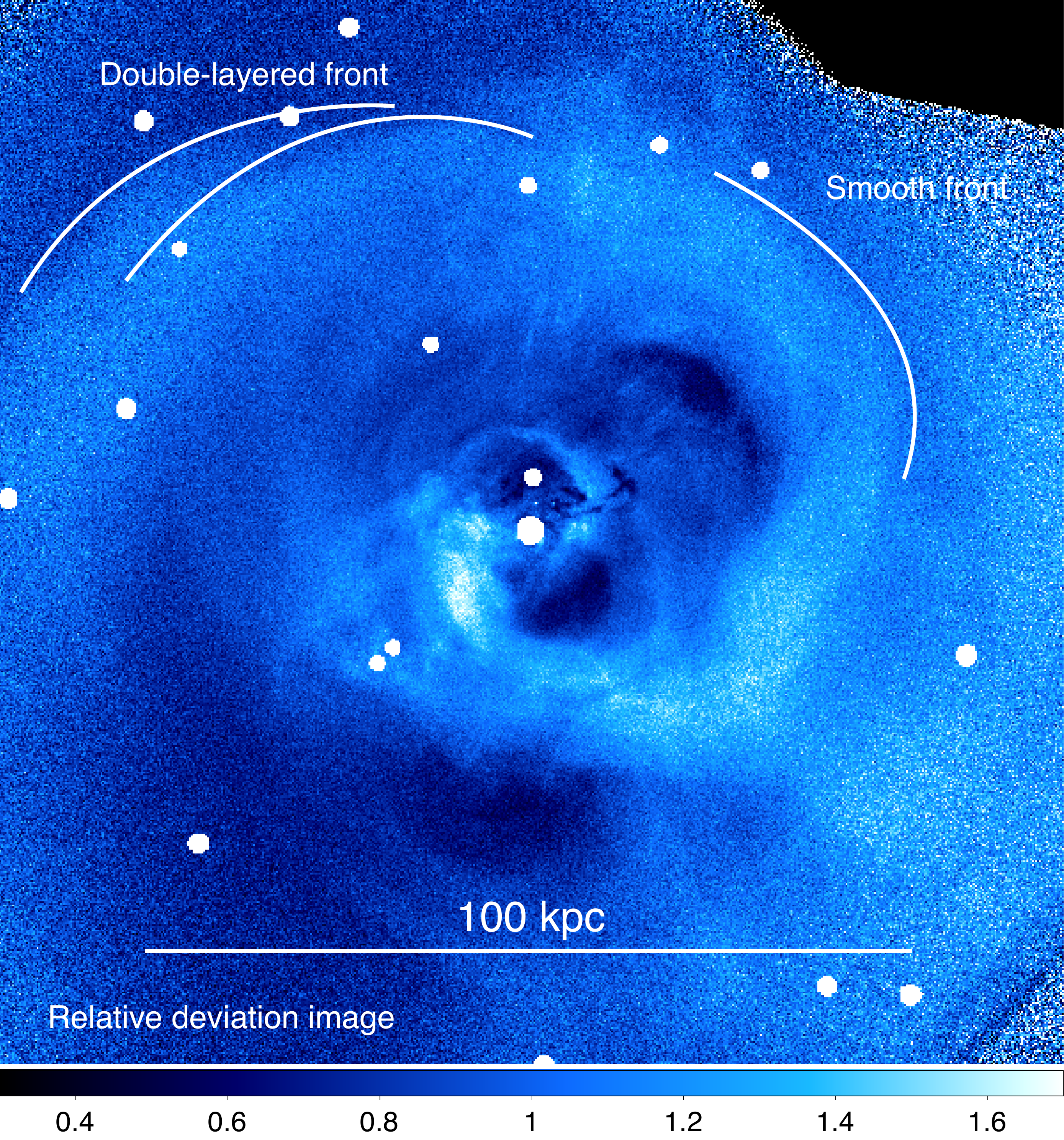}
 \end{minipage}
 \caption[]{{\it Left:} unsharp-masked image created by dividing the $\sigma=2$~pixel Gaussian smoothed fluximage by the $\sigma=20$~pixel Gaussian smoothed one. {\it Right:} relative deviation image with respect to the radial average.}
 \label{img:perseus_unsharp_residual}
\end{figure*}

We created the exposure and vignetting corrected {\it Chandra} images (fluximage) using the \verb+fluximage+ tool. We created an unsharp-masked image by dividing the $\sigma=2$~pixel Gaussian smoothed fluximage by the $\sigma=20$~pixel Gaussian smoothed one. The unsharp-masked image is shown in Figure~\ref{img:perseus_unsharp_residual} left. Also, in order to emphasize the low-contrast azimuthal variations, we divided the fluximage by the corresponding azimuthal average (with the centroid at the position of NGC~1275). The relative deviation image is shown in Figure~\ref{img:perseus_unsharp_residual} right.

\begin{figure*}
 \begin{minipage}{0.333\hsize}
  \centering
  \includegraphics[width=2.0in]{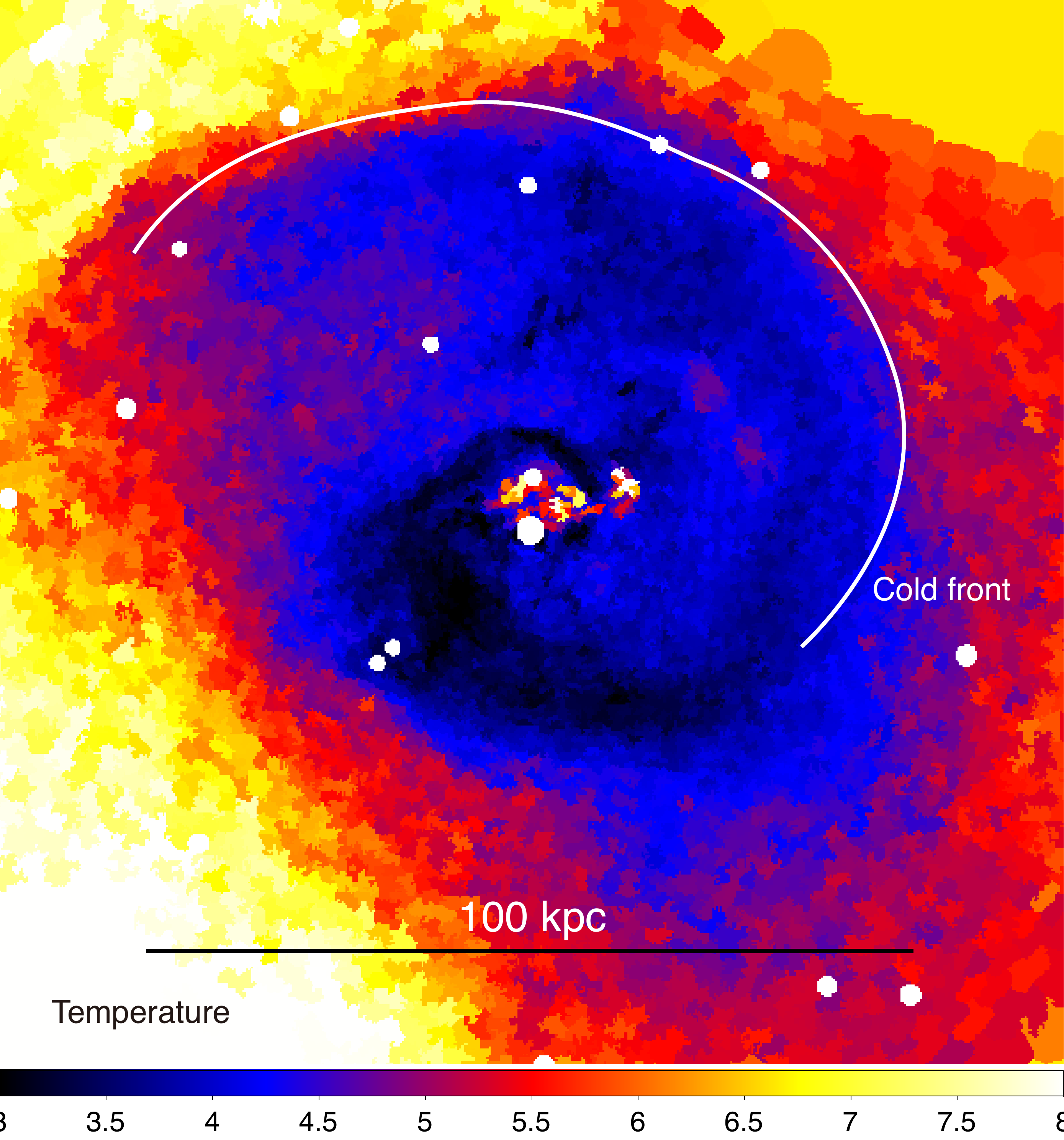}
 \end{minipage}%
 \begin{minipage}{0.333\hsize}
  \centering
  \includegraphics[width=2.0in]{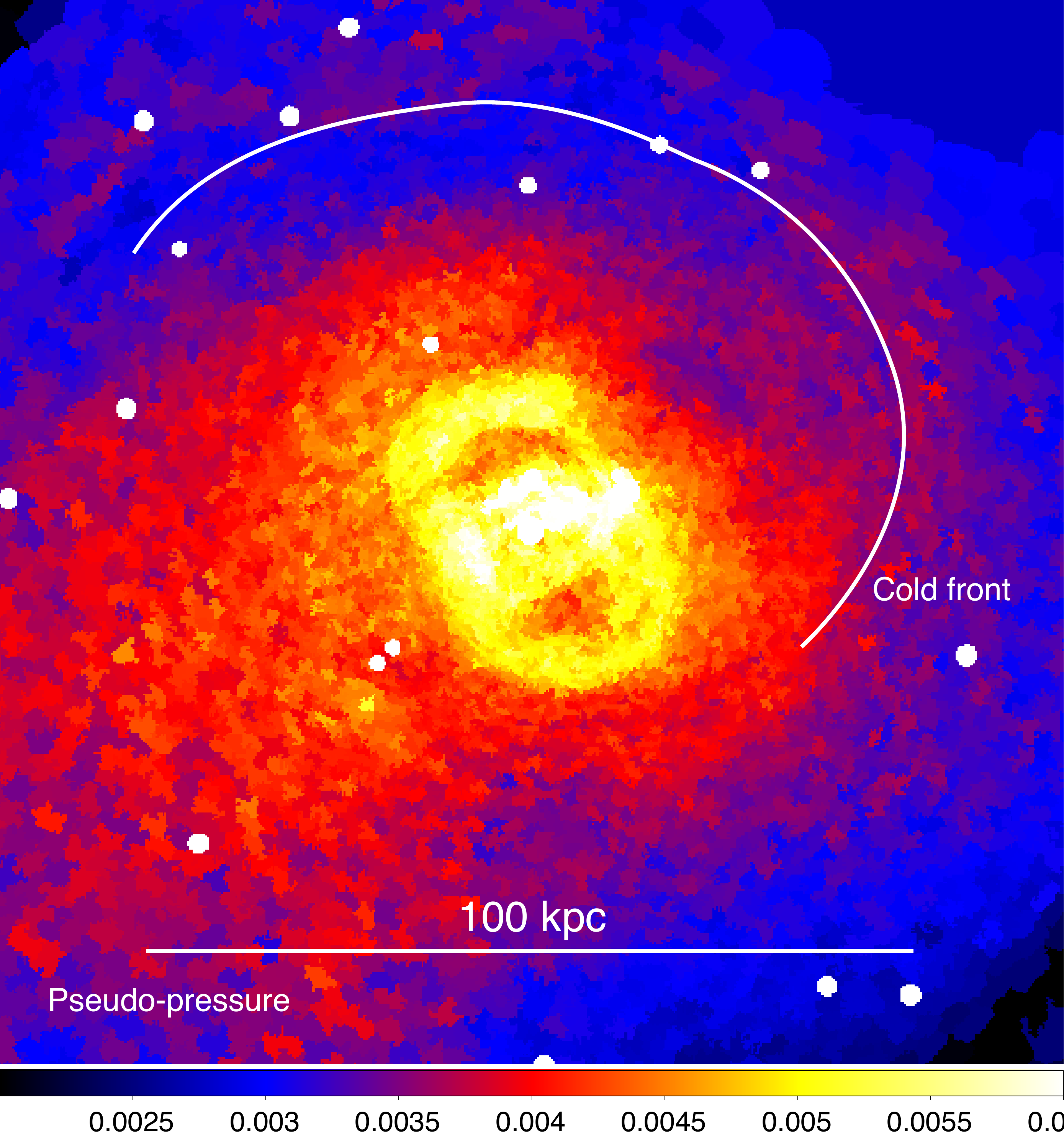}
 \end{minipage}%
 \begin{minipage}{0.333\hsize}
  \centering
  \includegraphics[width=2.0in]{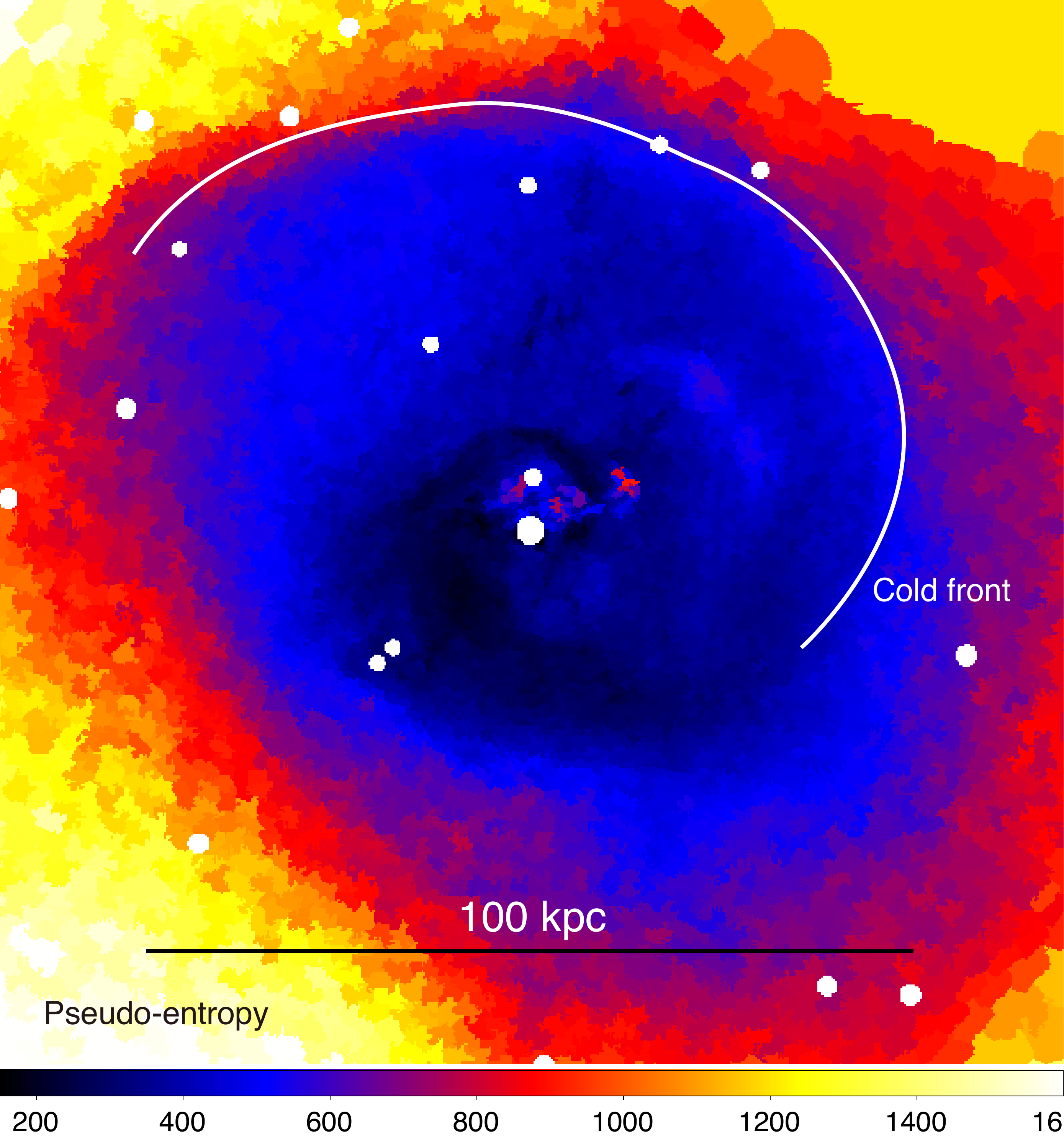}
 \end{minipage}
 \caption[]{Projected thermodynamic maps. {\it Left:} projected temperature map in the unit of keV. {\it Middle:} pseudo-pressure map. {\it Right:} pseudo-entropy map. The position of the cold front which delineates the spiral pattern shown in Figure~\ref{img:perseus_unsharp_residual} right is shown in white. The white circles are the positions of point sources which are visually identified and subtracted.}
 \label{img:perseus_thermo}
\end{figure*}

We also created the thermodynamic maps. We used the contour binning algorithm \citep{sanders06} to divide the field of view into small regions which are used for spectral fitting. The signal-to-noise ratio of each bin is about 100, corresponding to about 10000~counts/bin. We fitted the spectrum of each region using \verb+phabs(apec)+ model with the redshift fixed to 0.017284 and the hydrogen column density set to $1.38\times 10^{21}$~cm$^{-2}$, determined by the LAB (Leiden/Argentine/Bonn) radio HI survey \citep{kalberla05}. Using the best-fitting temperature $kT$ and normalization $\epsilon$, we calculated pseudo-pressure $\tilde{p}=kT\sqrt{\epsilon/A}$ and pseudo-entropy $\tilde{s}=kT(\epsilon/A)^{-1/3}$ where $A$ is the area of the corresponding region measured in the unit of pixels. The resulting maps are shown in Figure~\ref{img:perseus_thermo}.

\section{Results}
\subsection{Global morphological features}\label{sec:perseus_result_global}
In the unsharp-masked image (Figure~\ref{img:perseus_unsharp_residual} left) and the relative deviation image (Figure~\ref{img:perseus_unsharp_residual} right), we see plenty of structures. Since many of the features have already been mentioned and explored in the literature \citep[e.g.][]{churazov00,churazov03,fabian06,sanders07,sanders16b}, here we point out the structures which are related to our subsequent detailed analyses.

In the relative deviation image, we see a clear spiral-like pattern which has been discussed in the literature \citep[e.g.][]{churazov03,fabian06,sanders07,sanders16b}. Delineating the outer edge of the spiral, we see a brightness edge which starts about 50~kpc west of the core and extends anticlockwise to about 70~kpc north-east of the core in the unsharp-masked image. This spiral-like structure and the edge are also apparent in the projected temperature and pseudo-entropy maps (Figure~\ref{img:perseus_thermo}). The fact that the gas beneath the edge is cooler and has lower-entropy than above the edge, as well as the fact that no clear pressure structure along the edge is seen, indicate that the edge is a cold front, originating from the sloshing motion of cool gas in the core \citep{ascasibar06,markevitch07}.

In addition to known structure, we identify two new structures; (1) The west half of the front seems relatively smooth, while the east half of the front exhibits a more complex, double-layered structure, indicating the existence of developing instability which may be due to the sloshing motion of the ICM. We point out the similarity between this cold front and the low-viscosity numerical simulation result by \citet{roediger13a} \citep[see the middle panel of Figure~3 of][]{roediger13a}. They attributed this phenomenon to the difference of mean shear at different azimuths of the front; (2) In the unsharp-masked image, underneath the west half of the front, we see feather-like structures, namely alternating bright and faint regions which have not been reported previously in the literature. Recently \citet{werner15} found similar structures just below the northwestern cold front in the Virgo cluster. \citet{wang18} also reported similar narrow ``channels'' in the surface brightness image of Abell~2142.\\

\subsection{Double-layered structure of the eastern cold front}\label{sec:perseus_result_eastcf}
\begin{figure*}
 \begin{minipage}{0.495\hsize}
  \centering
  \includegraphics[width=3.0in]{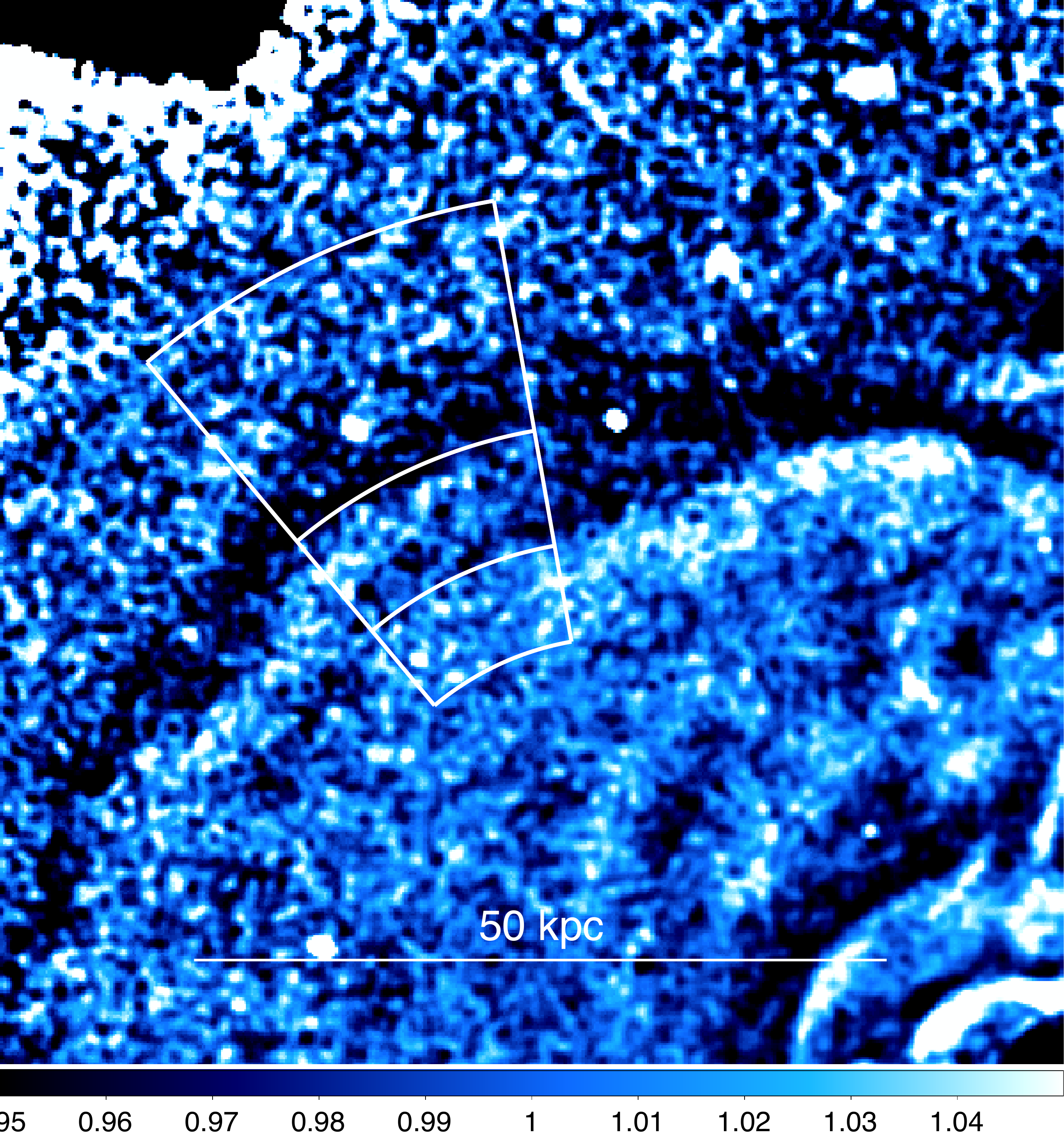}
 \end{minipage}
 \begin{minipage}{0.495\hsize}
  \centering
  \includegraphics[width=3.0in]{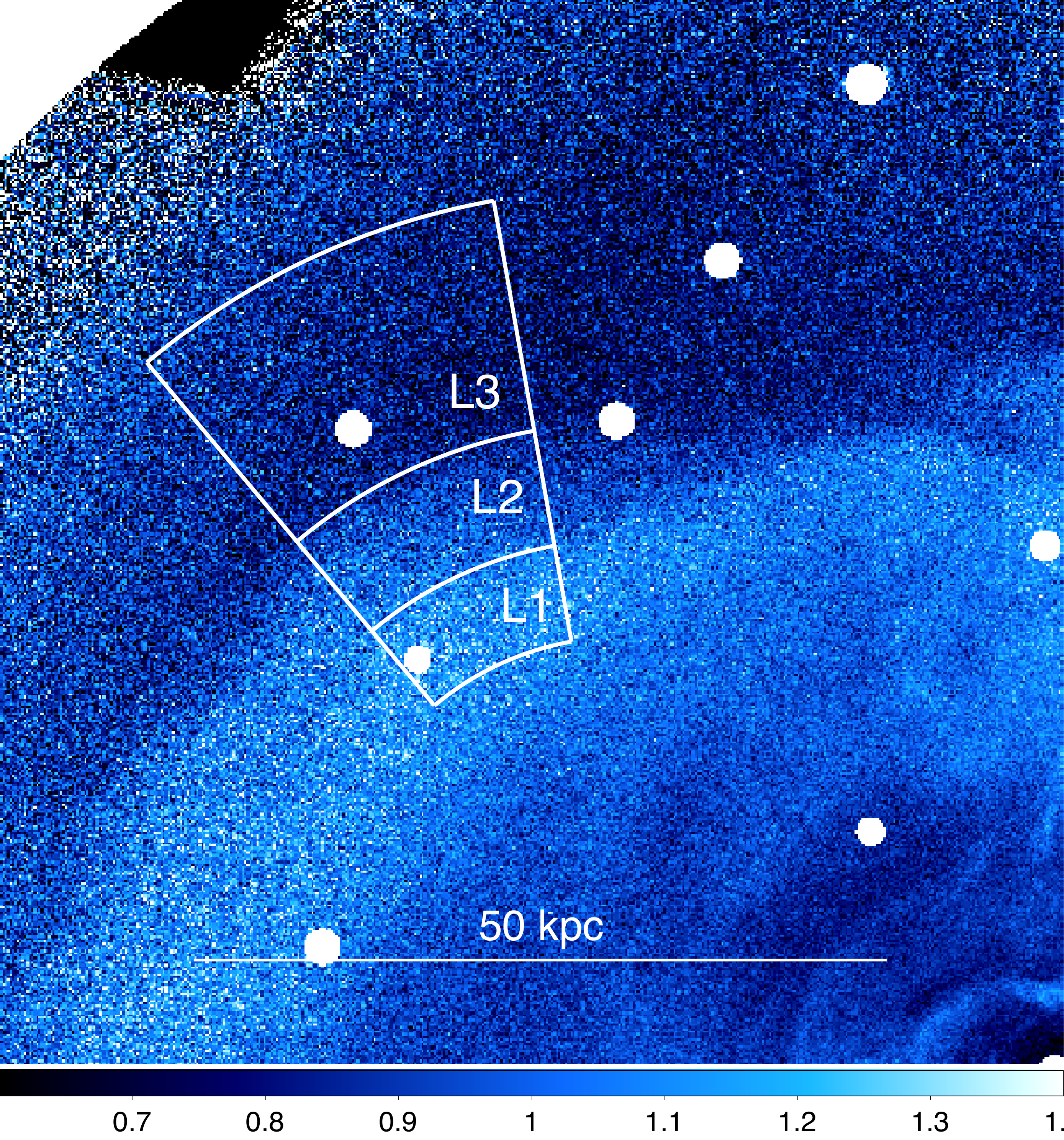}
 \end{minipage}
 \caption[]{Same as Figure~\ref{img:perseus_unsharp_residual}, zoomed in on the double-layered structure. White partial annuli denote the regions along which the surface brightness profile and X-ray spectra are extracted.}
 \label{img:perseus_unsharp_residual_eastcf}
\end{figure*}
Figure~\ref{img:perseus_unsharp_residual_eastcf} shows the close-up view of the eastern part of the cold front, where we see a peculiar double-layered structure. The overlaid partial annuli are adjusted so that their curvature matches the curvatures of the two fronts.

\begin{figure*}
  \centering
  \includegraphics[width=7.0in]{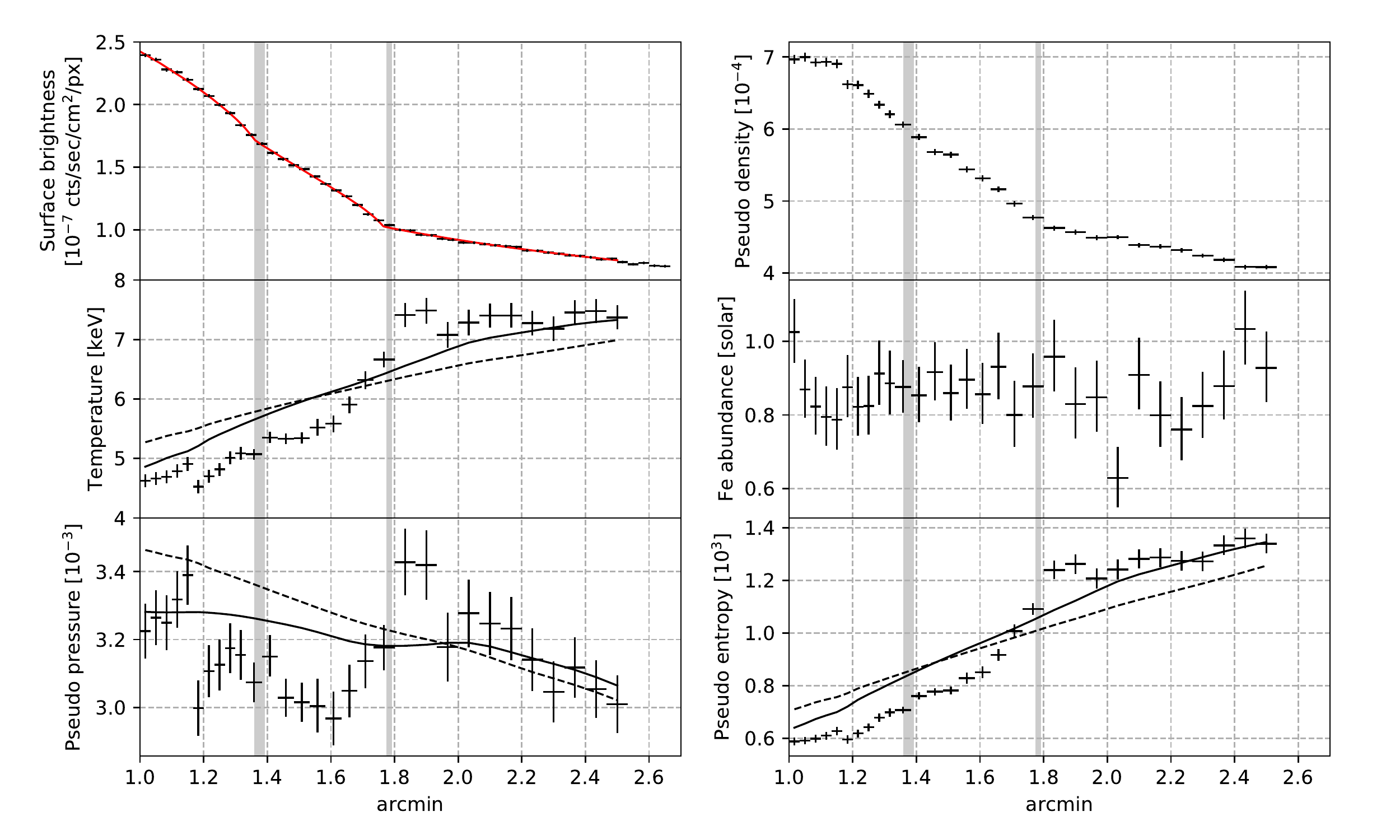}
 \caption[]{{\it Top left:} surface brightness profile extracted across the double-layered structure. The red curve is the best-fitting projected double-broken power-law model. {\it Top right:} pseudo-density profile. {\it Middle left:} projected temperature profile. {\it Middle right:} projected Fe abundance profile. {\it Bottom left:} pseudo-pressure profile. {\it Bottom right:} pseudo-entropy profile. The solid/dashed curves are reference profiles calculated using the azimuthally averaged profile over 60$^\circ$-180$^\circ$/the entire azimuths. The gray vertical bands denote the positions of the breaks in the surface brightness profile. The radial distance from the centre at which all the plots start is $\sim$2.5~amin.}
\label{img:perseus_prof_eastcf}
\end{figure*}
The extracted surface brightness profile is shown in Figure~\ref{img:perseus_prof_eastcf} top left. In order to estimate the positions of the breaks, we fitted the surface brightness profile using a projected double-broken power law model following the procedure presented by \citet{ichinohe17}. The centre of spherical symmetry is set to the concentre of the sectors shown in Figure~\ref{img:perseus_unsharp_residual_eastcf} left. The model describes the profile well with $\chi^2/\mr{DOF}=40.7/37$ and strongly prefers the double-broken power law model to the single-broken power law model by $\Delta\chi^2/\Delta\mr{DOF}=104.5/3$. The positions of the two breaks are $r_{12}=82.5\pm1.0$~arcsec and $r_{23}=107.0\pm0.5$~arcsec respectively for the inner and outer breaks. These error ranges are overlaid in Figure~\ref{img:perseus_prof_eastcf} as the grey vertical bands.

We also extracted thermodynamic profiles along the sectors shown in Figure~\ref{img:perseus_unsharp_residual_eastcf} left. The spectral fitting was performed in the same manner as was done for the thermodynamic maps (Figure~\ref{img:perseus_thermo}). The resulting (pseudo-) thermodynamic profiles are shown in Figure~\ref{img:perseus_prof_eastcf}.

We also plotted reference profiles in the temperature, the pressure and the entropy panels, using the azimuthally averaged temperature, pseudo-pressure and pseudo-entropy profiles. We split the entire field of view into an annular grid with radial and azimuthal intervals of 10~arcsec and 10$^\circ$ respectively, which are concentred on the cluster centre. From each grid element, we extracted spectra and analyzed them in the same way mentioned in the paragraphs above, resulting in thirty-six thermodynamic values per radial annulus. We averaged them to obtain the radial thermodynamic profiles. Since the projected pressure map shows a significant asymmetry as shown in Figure~\ref{img:perseus_thermo}, we calculated the averaged thermodynamic profiles in two ways; (1) averaged over all the azimuths and (2) averaged over the azimuthal range of 60$^\circ$-180$^\circ$. The reference profiles based on the average over all azimuths are plotted using the dashed curves and the ones using the azimuthal range of 60$^\circ$-180$^\circ$ are plotted using the solid curves in Figure~\ref{img:perseus_prof_eastcf}.

Although almost continuous, we see some indications for the change in slope and the existence of a mild jump around the first break in both the temperature and entropy profiles. These profiles show a rapid increase toward the second break, and flatten out beyond the second break. These profiles are systematically lower than the reference profiles and seem to overtake the reference profiles around the second break. While the pressure profile is continuous at the first break, it shows 10--15\% jump at just outside the second break and monotonically decreases beyond the second break; the overall pressure profile is non-monotonic and a dip is seen between the two breaks. The Fe abundance is almost constant over the entire radial range.

\subsection{Feather-like structures}\label{sec:perseus_feather_morphology}
\begin{figure*}
 \begin{minipage}{0.333\hsize}
  \centering
  \includegraphics[width=2.0in]{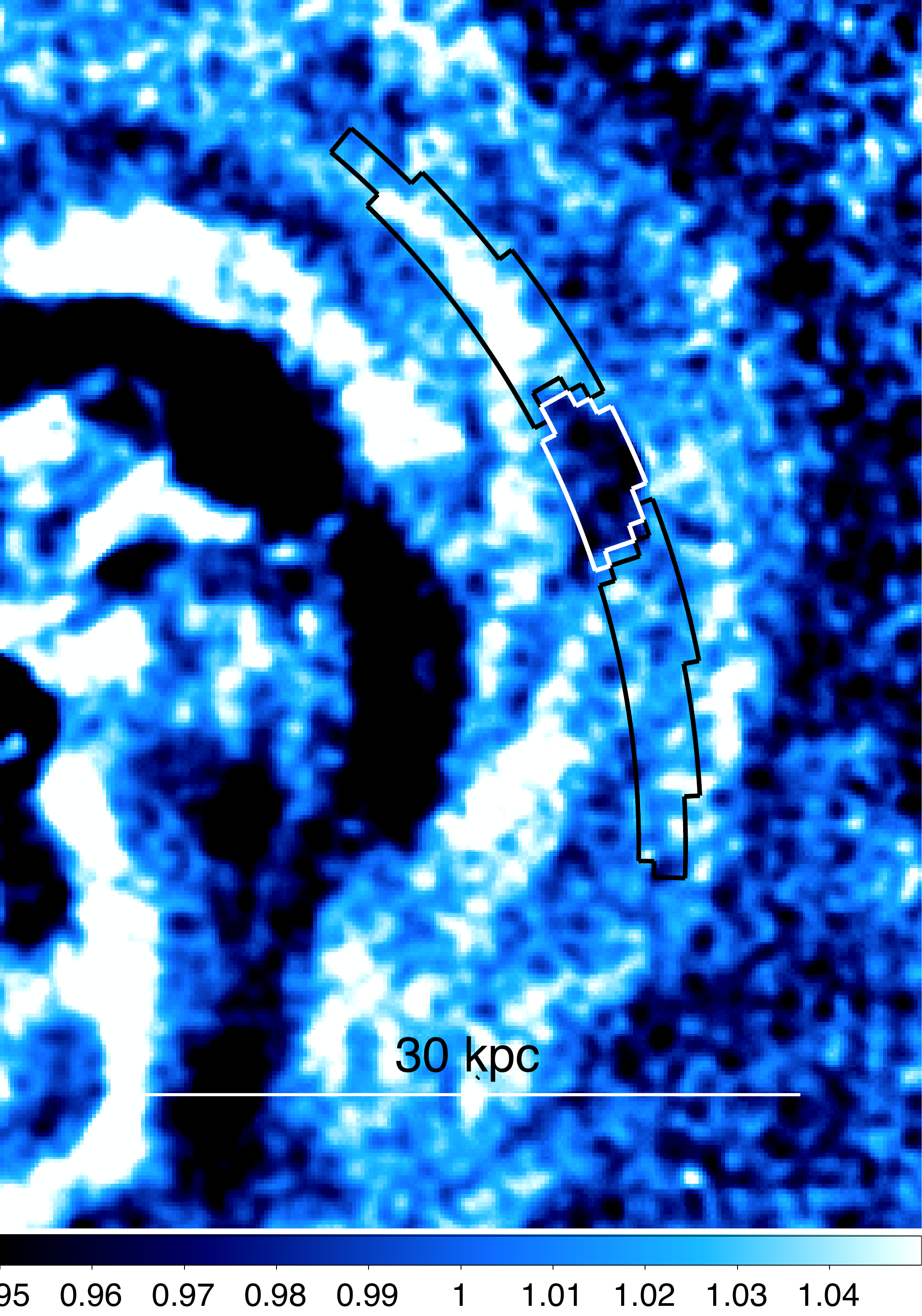}
 \end{minipage}%
 \begin{minipage}{0.333\hsize}
  \centering
  \includegraphics[width=2.0in]{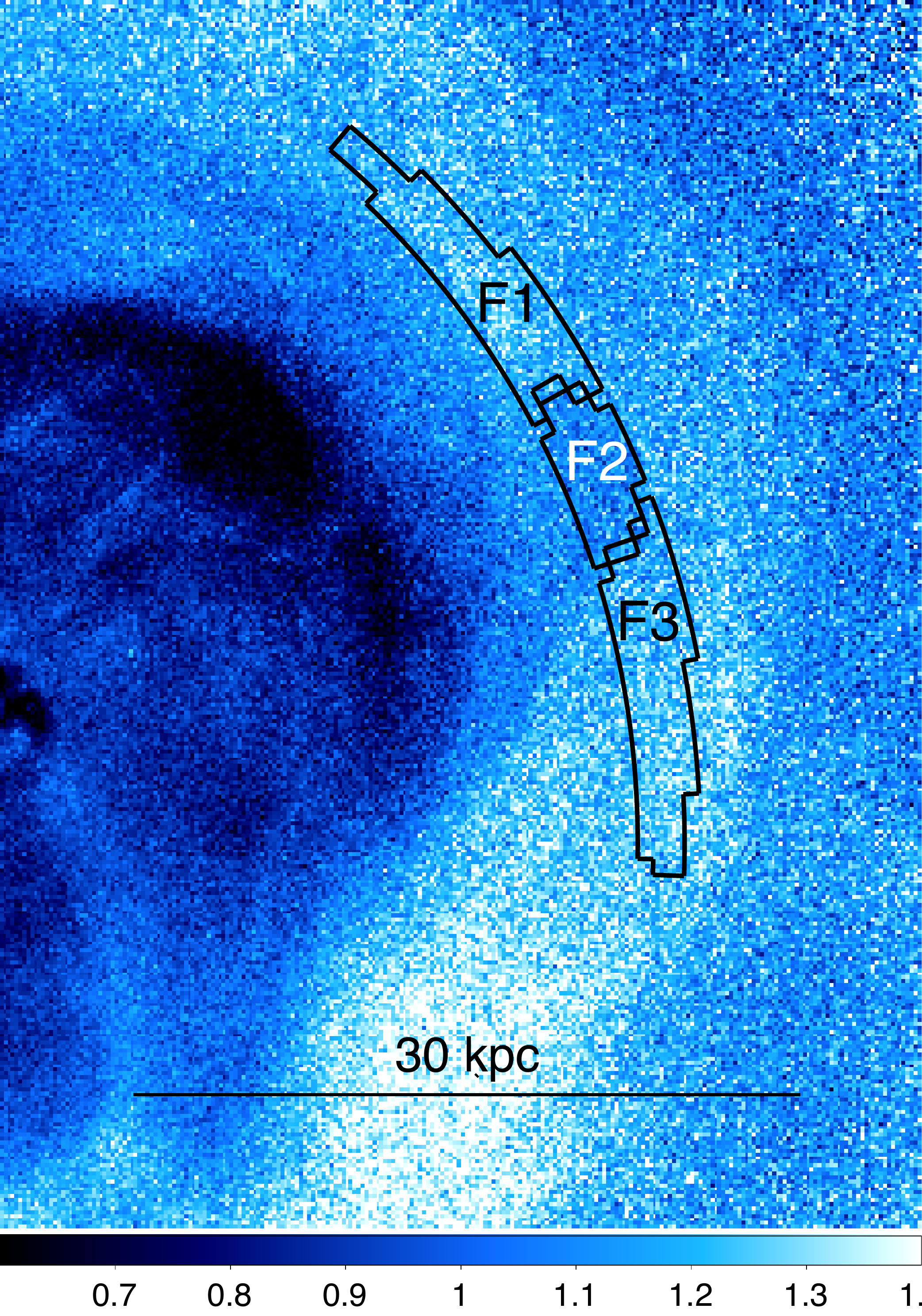}
 \end{minipage}%
 \begin{minipage}{0.333\hsize}
  \centering
  \includegraphics[width=2.0in]{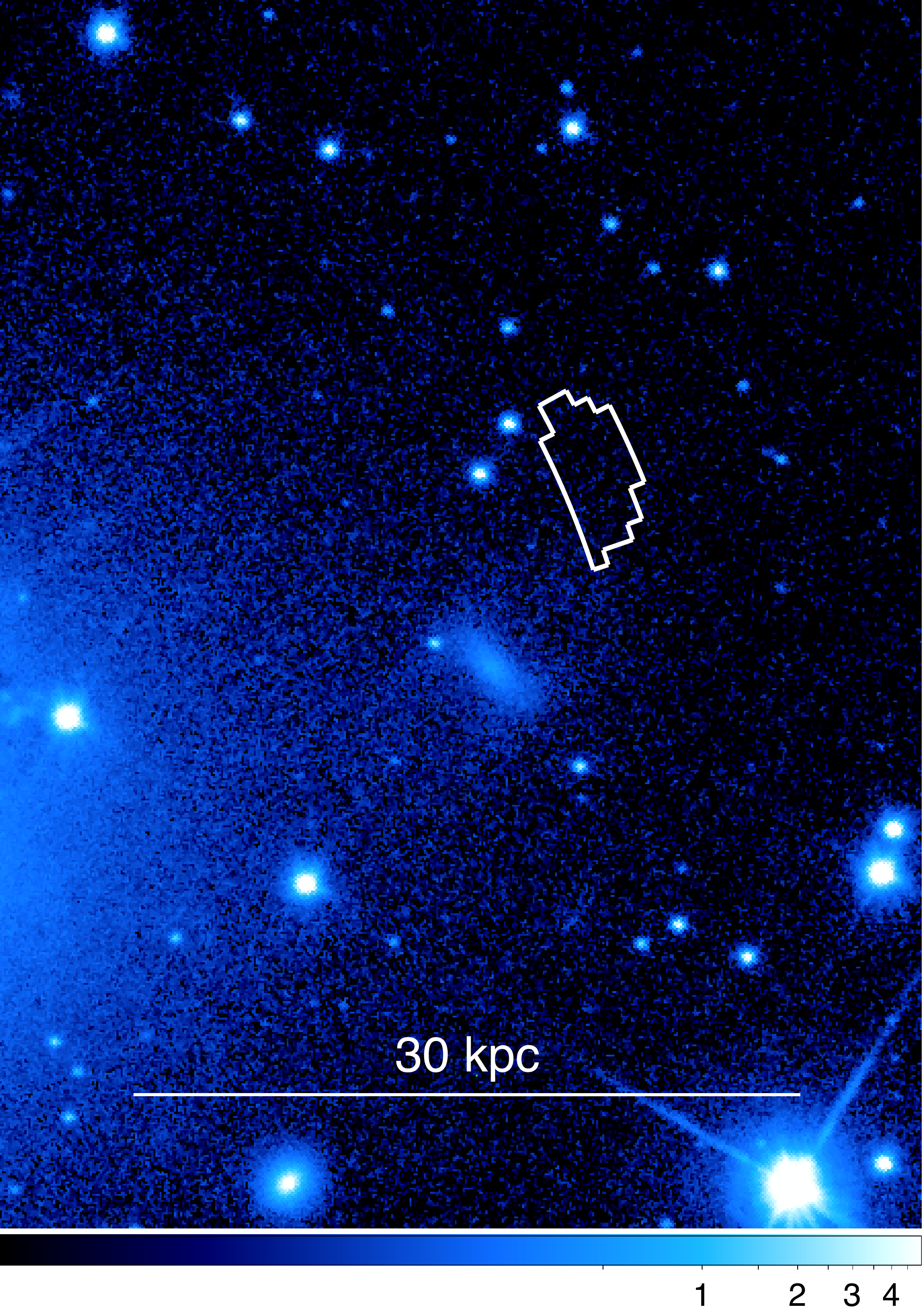}
 \end{minipage}
 \caption[]{The closeup view of the unsharp-masked image (left), the relative deviation image (middle) and the SDSS $r$-band optical image of the corresponding sky region. The regions employed in a detailed study of the brightness dip are overlaid on the relative deviation image. These regions are numbered from F1 to F3 from north to south.}
\label{img:perseus_unsharp_residual_zoom}
\end{figure*}

Figure~\ref{img:perseus_unsharp_residual_zoom} shows the same images as Figure~\ref{img:perseus_unsharp_residual}, zoomed-in to the vicinity of the feather-like structures, together with the SDSS $r$-band optical image \citep{eisenstein11,ahn14}. The alternating bright and faint regions are clearly seen in the unsharp-masked image (left panel). Among the structures which are apparent in the unsharp-masked image, the central faint region ({\it brightness dip}; denoted by the white polygon in the leftmost panel) exhibits the most prominent contrast against the surrounding ICM, which even can be seen in the lower-contrast relative deviation image (the polygon with the annotation ``F2'' in the middle panel).

To investigate the ICM properties in the brightness dip in detail, we chose the regions based on the relative deviation image, because the unsharp-masked image is essentially an edge-enhanced image and thus is not suitable for considering the absolute brightness of the structures. We number these regions from F1 to F3 from north to south (see Figure~\ref{img:perseus_unsharp_residual_zoom} middle). The areas of the regions~F1 and F3 are double compared to that of the region~F2 for each radius, lessening the uncertainty due to the difference of the radial dependency of the region shape between the regions.

\begin{table*}
 \centering
 \cprotect\caption{Best-fitting parameters of the single temperature model in the ``feathers'' region. (a) Calculated using $\epsilon/S$ where $\epsilon$ is \verb+apec+ normalization and $S$ is the region area in arcsec$^2$.}
 \begin{tabular}{c|cccc}
  \hline
Region & Temperature (keV) & Fe abundance (solar) & Area-normalized \verb+apec+ norm$^\mr{a}$ (10$^{-6}$) & nH ($10^{22}$~cm$^{-2}$)\\
  \hline
 F1 & 3.81$\pm$0.03 & 0.85$\pm$0.02 & 3.49$\pm$0.03 & 0.154$\pm$0.002 \\
 F2 & 3.89$\pm$0.05 & 0.88$\pm$0.04 & 3.15$\pm$0.04 & 0.151$\pm$0.004 \\
 F3 & 3.96$\pm$0.03 & 0.86$\pm$0.02 & 3.53$\pm$0.03 & 0.149$\pm$0.002 \\
  \hline
 \end{tabular}
 \label{tbl:perseus_prof_dip_1t}
\end{table*}

For each region, we extracted spectra and fitted the data with an absorbed single-temperature thermal plasma model; \verb+phabs(apec)+. We modeled each region independently allowing all the parameters to vary except for the redshift. The resulting parameters are shown in Table~\ref{tbl:perseus_prof_dip_1t}. Although the normalization shows a clear deficit with respect to the surroundings in the brightness dip region (region~F2), we do not see significant deviations of the other quantities. Note that we also tried to model these regions using two temperature model, but the parameters were not well constrained.

\section{Discussion}
\subsection{Double-layered structure}\label{sec:perseus_eastcf}
As shown in Section~\ref{sec:perseus_result_eastcf}, the projected double-broken power law model represents the surface brightness profile around the eastern part of the cold front (Figure~\ref{img:perseus_unsharp_residual_eastcf}) better than the single-broken power law model does. Although the best-fitting parameters other than the break radii are uncertain because the assumption of the spherical symmetry in the projection is probably inaccurate, at least qualitatively the underlying density profile must then also host a double-layered structure similar to the surface brightness profile or the images.

\subsubsection{Kelvin-Helmholtz instability}
Generally, at a shock front, when the density shows a jump, the temperature, pressure and entropy also show a jump, because a shock front propagating through the ICM heats and compresses the gas behind it. In contrast, since cold fronts are merely the interface between a cold gas parcel and hot ambient medium, when the density exhibits a drop, the temperature and entropy exhibit a jump, resulting in an almost continuous pressure profile across the front.

As shown in Figure~\ref{img:perseus_prof_eastcf}, both at the first and the second breaks, the density decreases but the temperature and entropy increase, which indicates that neither feature is a shock front. Especially, despite the jumps of temperature and entropy, the pressure is almost continuous at the first break, suggesting that this represents a (mild) cold front. At the second break, on the other hand, the change of slopes (rapid increase to flat) of the temperature and entropy profiles is associated also with a jump in the pressure, suggesting this structure is not a cold front in the classical sense.

\begin{figure}
  \centering
 \includegraphics[width=2.5in]{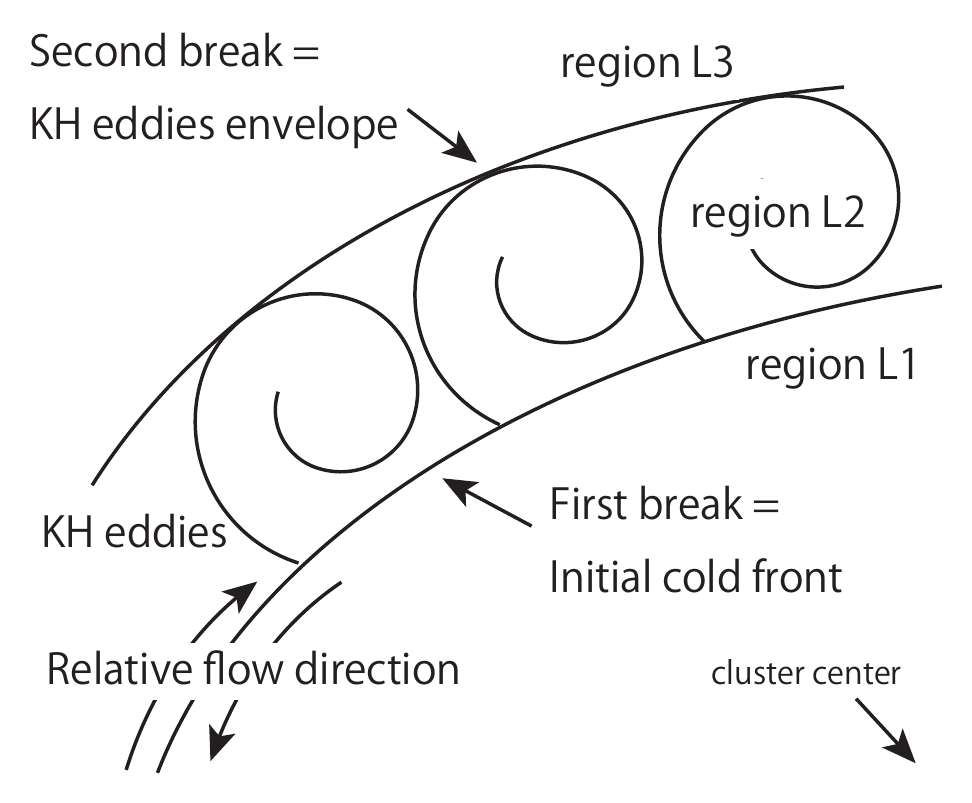}
 \hspace{5mm}
 \caption[]{Schematic illustration of the double-layered structure due to the developing KHIs on top of the sloshing cold front. The region numbers correspond to the ones in Figure~\ref{img:perseus_unsharp_residual_eastcf} right. Note that the outwards motion of the cold front is not represented in this simplified cartoon.}
\label{img:perseus_khi_schem}
\end{figure}
 Such a situation can be realized when Kelvin-Helmholtz instabilities (KHIs) are developing on top of the cold front: the first break was originally a cold front due to the sloshing motion of the gas induced by a previous merger event, which currently manifests itself with the spiral-shaped morphology. Due to the shearing motion of the gas, KHIs have been set off and are currently developing, and the second break represents the envelope of current maximum heights of the KH eddies. Since the KHIs mix the gas inside the front into the ambient gas, the contrast of the initial cold front is being weakened, and the thermodynamic properties inside and outside the first break are not strongly discontinuous. Since KHI eddies are not as coherent as the cold front, we would not detect a clear jump of temperature or entropy at the second break but would observe continuous changes of thermodynamic profiles, which is what our results actually show. A simplified cartoon of this situation is schematically drawn in Figure~\ref{img:perseus_khi_schem}. It should be noted that the sloshing cold front is itself a wave phenomenon that propagates outwards in the stratified atmosphere, therefore the fluid structure at the interface is in fact different than the shear interface realised in simple, two-dimensional parallel-flow setups. 3D numerical simulations of gas sloshing demonstrate that KHI with properties very similar to our observations develop even in the conditions of this more complex flow pattern \citep{roediger13b}.

In order to test this scenario, we examined the multi-temperature properties of the corresponding regions. We combined the sectors below the inner break, between the two breaks, and above the outer break (regions~L1, L2 and L3, respectively). These regions are shown in Figure~\ref{img:perseus_unsharp_residual_eastcf}. We extracted spectra using these regions and fitted them using single-temperature (1T; \verb|phabs(apec)|) and two-temperature (2T; \verb|phabs(apec+apec)|) models. Note that three-temperature modeling does not improve the fit. In addition to the independent fitting, we also model the spectra in all the regions simultaneously, with the two temperatures, Fe abundance and $n_H$ tied together.

\begin{table*}
 \centering
 \caption{Summary of the two-temperature fitting results for the double-layered structure. $(a)$ APEC normalization; $\epsilon = 10^{-14}\int n_e n_H dV/4\pi[D_A(1+z)^2]$ where $D_A$ is the angular diameter distance to the source (cm), $n_e$ and $n_H$ are the electron and hydrogen ion densities (cm$^{-3}$)}
 \begin{tabular}{c|cccc}
 \hline\hline
  Fitting condition & $kT_1$ (keV) & $\epsilon_1$ $(10^{-3})^\mr{a}$ & $kT_2$ (keV) & $\epsilon_2$ $(10^{-3})^\mr{a}$\\
 \hline
 1T, region~L1           & $4.53\pm0.04$          & $1.36\pm0.01$          & --                     & --                     \\
 1T, region~L2           & $5.21\pm0.04$          & $1.50\pm0.01$          & --                     & --                     \\
 1T, region~L3           & $6.75\pm0.05$          & $2.57\pm0.01$          & --                     & --                     \\
 2T, region~L1           & $6.36_{-0.42}^{+0.52}$ & $0.92\pm0.10$          & $2.62_{-0.20}^{+0.17}$ & $0.47\pm0.10$          \\
 2T, region~L2           & $9.03_{-1.18}^{+0.98}$ & $0.80_{-0.32}^{+0.17}$ & $3.40_{-0.32}^{+0.39}$ & $0.72_{-0.17}^{+0.19}$ \\
 2T, region~L3           & $8.34_{-0.28}^{+0.97}$ & $2.21_{-0.34}^{+0.09}$ & $2.94_{-0.32}^{+0.96}$ & $0.37_{-0.09}^{+0.29}$ \\
 2T, region~L1 (comb)    & $8.37_{-0.24}^{+0.27}$ & $0.59\pm0.05$          & $3.23_{-0.11}^{+0.09}$ & $0.78\pm0.05$          \\
 2T, region~L2 (comb)    & --                     & $0.89\pm0.05$          & --                     & $0.63\pm0.05$          \\
 2T, region~L3 (comb)    & --                     & $2.15_{-0.07}^{+0.06}$ & --                     & $0.44_{-0.06}^{+0.07}$ \\
 \hline\hline
 & Fe abundance (Solar) & $n_H$ ($10^{22}~\mr{cm}^2$) & Cstat/DOF & \\
 \hline
 1T, region~L1           & $0.78_{-0.02}^{+0.03}$ & $0.160\pm0.002$ & 624.8/467 & \\
 1T, region~L2           & $0.79\pm0.03$          & $0.160\pm0.002$ & 730.8/467 & \\
 1T, region~L3           & $0.77\pm0.02$          & $0.158\pm0.002$ & 911.6/467 & \\
 2T, region~L1           & $0.72\pm0.03$          & $0.163\pm0.003$ & 533.4/465 & \\
 2T, region~L2           & $0.81\pm0.03$          & $0.159\pm0.003$ & 641.0/465 & \\
 2T, region~L3           & $0.81\pm0.03$          & $0.157\pm0.002$ & 829.0/465 & \\
 2T, region~L1 (comb)    & $0.79\pm0.02$          & $0.159\pm0.001$ & 2014.1/1403 & \\
 2T, region~L2 (comb)    & --                     & --              & --          & \\
 2T, region~L3 (comb)    & --                     & --              & --          & \\
 \hline\hline
 \end{tabular}
 \label{tbl:perseus_pdip}
\end{table*}

Table~\ref{tbl:perseus_pdip} shows the fitting results. While every region prefers the 2T modelling ($\Delta$Cstat/$\Delta\mr{DOF}$=91.4/2, 89.8/2 and 82.6/2 for regions~L1, L2 and L3, respectively), letting the temperature, Fe abundance and $n_H$ in all the regions vary independently does not significantly improve the fitting ($\Delta$Cstat/$\Delta\mr{DOF}$=10.79/8), which suggests all the regions share the same two temperatures. Moreover, the ratio of the normalization of the cooler component to that of the hotter component ($\epsilon_2/\epsilon_1$) increases toward the cluster centre (regions~L3 to L1). This supports the interpretation of the double-layered structure being a developing KHI, where the hotter component represents the ambient medium, while the cooler component represents the gas which is originally inside the cold front, being mixed into the ambient medium via the currently developing KHI.

Using numerical simulations, \citet{roediger13a} indeed showed that KHIs can be induced along the edge of the sloshing spiral under certain conditions (particularly in low-viscosity cases), and that in such cases, the surface brightness profile hosts characteristic multiple edges, similarly to our case. It is also suggested by \citet{roediger13a,roediger13b} that the distance between the edges is about a fourth to a half of the scale length of the KH rolls. In our case, the distance between the two breaks is $25\pm 1$~arcsec, corresponding to the actual distance of $8.8\pm0.4$~kpc, while the azimuthal extension of the double-layered structure is 30-40~kpc, which is consistent with the prediction from the simulation. Therefore, we suggest that this double-layered structure originates from the sloshing cold front accompanied by developing KHIs.

Our detection joins a growing number of indications for the existence of KHI suggested recently. For example, a similar multiple-edge structure in the merger cold front of the NGC~1404 galaxy has been reported by \citet{su17}, and the existence of both the multiple-edge structure and the possible pressure deficit (see also the next section) has been presented also for the merger cold front in Abell~3667 \citep{ichinohe17}. \citet{werner16} found multiple sloshing-induced cold fronts in the core of the Ophiuchus cluster and proposed that they could be due to KHIs. \citet{walker17} has suggested the existence of another giant KHI roll at $\sim$150~kpc southeast from the core in association with the outer cold front of the Perseus cluster. Recently \citet{wang18} found well-developed KH eddies in the southern cold front of Abell~2142.

\subsubsection{Pressure deficit and nonthermal pressure support}
Based on the KHI scenario, here we investigate the ICM microphysical properties. As is pointed out in Section~\ref{sec:perseus_result_eastcf}, the pressure between the two breaks seems to be insufficient to balance the surrounding gas. Assuming that the gas is uniform over the line-of-sight depth of $L$, the pseudo-density $\tilde{n}$ is translated to the electron density value of $n_e = 38\tilde{n}(L/63~\mr{kpc})^{-1/2}\mr{cm}^{-3}$, where 63~kpc corresponds to the approximate distance of the structure from the cluster centre ($\sim$3~arcmin). Similarly, the physical value of the pressure deficit $\Delta p$ corresponding to the deficit of pseudo-pressure $\Delta\tilde{p}\sim0.1\times10^{-3}$ is $\Delta p_e\sim 3.8\times10^{-3}~\mr{keV~cm^{-3}}(L/63~\mr{kpc})^{-1/2}$.

In the case of the two-temperature modeling (see Table~\ref{tbl:perseus_pdip}), assuming that the cooler component of the gas inside region~L1 is uniform over the line-of-sight depth of $L_1$, the normalization of the \verb+apec+ $\epsilon_1$ is translated to the physical density value of $n_1 \sim 1.4\times 10^{-2} (L_1/63~\mr{kpc})^{-1/2}\mr{cm}^{-3}$. Similarly, that of the hotter component $\epsilon_2$ is translated to $n_2\sim7.0\times 10^{-3} (L_2/334~\mr{kpc})^{-1/2}\mr{cm}^{-3}$, where 334~kpc corresponds to the core radius $r_c = 15.85~\mr{arcmin}$ of the best-fitting $\beta$-model for the northeastern direction obtained by \citet{urban14}. Accordingly, the pressure of each component is estimated at $p_1\sim4.4\times10^{-2}(L_1/63~\mr{kpc})^{-1/2}\mr{keV~cm}^{-3}$ and $p_2\sim5.9\times10^{-2}(L_2/334~\mr{kpc})^{-1/2}\mr{keV~cm}^{-3}$.

Although there are geometrical uncertainties, in both cases, the electron pressures seem to disagree with each other by a value of order $10^{-2}~\mr{keV~cm^{-3}}$, with the gas between the two breaks or the cooler component having a lower pressure.

There are some candidates which can support the pressure deficit. The first one is magnetic pressure \citep[e.g.][]{keshet10}. In this case, assuming that the pressure deficit is fully supported by the magnetic pressure, $B^2/8\pi =\Delta p$ yields the magnetic field strength of $B\sim30~\mu\mr{G}$, where $p = (n_e+n_i)kT$ is the pressure, assuming equal temperature between electrons and ions, and $n_e=1.2n_i$. \citet{reiss14} found a thermal pressure jump of $\sim$10\% in a sample of 17 cold fronts in relaxed clusters not including Perseus. If the Perseus cold front is in fact located at the second break, our estimated pressure deficit of a similar magnitude is consistent with their scenario. However, our interpretation is that the real location of the cold front corresponds to the first break in the surface brightness profile instead.

As gas sloshing induces gas motion, turbulent pressure can be another candidate. In this case, assuming the turbulence to be isotropic, $\rho V_\mr{1d}^2 =\Delta p$ leads to the one-component turbulent strength of $V_\mr{1d}\sim400~\mr{km~s^{-1}}$, where the mass density $\rho=\mu m_p n_e$ with $m_p$ being the proton mass and $\mu=0.6$ being the mean particle weight.

The forming eddies themselves can also contribute to the pressure deficit because the central part of a vortex has lower pressure than its outer parts. This mechanism can have a significant impact especially in the early stage of the development of KHIs when the eddies remain coherent. It is difficult to distinguish these two mechanisms (turblence and eddy) from the current observation. Direct measurement of the emission line width with future calorimeter observations of this region using e.g. {\it XRISM} \citep{tashiro18} and {\it Athena} \citep{barcons15} would be thus interesting regarding the origin of this feature and also more generally the pressure component at work in the cores of galaxy clusters.

KH eddies collapse into smaller-scale eddies and ultimately dissipate into heat, and the gas will be turbulent during the dissipation. Therefore, it is interesting to estimate whether or not the turbulent heating can balance the radiative cooling at this radius, assuming that the current size of the structure and the current turbulent strength represent the driving scales of turbulence. According to \citet{zhuravleva14}, the turbulent heating rate $Q_\mr{turb}$ can be estimated using $Q_\mr{turb}=C_\mr{Q}\rho V_\mr{1d}^3/l$, where $C_Q\sim 5$ is a fiducial constant related to the Kolmogorov constant and may differ by a factor $\sim 2$, $V_\mr{1d}$ is the one-component velocity of the turbulence, and $l$ is the corresponding spatial scale. Using $l=30-40~\mr{kpc}$ and $V_\mr{1d}\sim400~\mr{km~s^{-1}}$, the heat input rate can be estimated at $Q_\mr{turb}\sim3\times10^{-26}~\mr{erg~cm^{-3}~s^{-1}}$.

On the other hand, the cooling rate of the gas $Q_\mr{cool} = \Lambda n_e n_i$ is estimated at $1-2\times10^{-27}\mr{erg~cm^{-3}~s^{-1}}$, where $\Lambda$ is the normalized cooling function calculated in \citet{sutherland93}, with the temperature of several keV and approximating the ICM as having a Solar metal abundance. Considering the uncertainties (e.g., geometry and the value of $C_Q$), we can only suggest that this cooling rate is comparable to $\lesssim Q_\mr{turb}$, meaning that the turbulent heating is able to balance the radiative cooling.

The magnetic pressure support and the turbulent pressure support can coexist, and it is difficult to disentangle the two factors. However, it is worth pointing out that our estimated value of $Q_\mr{turb}\sim3\times10^{-26}~\mr{erg~cm^{-3}~s^{-1}}$ agrees within an order of magnitude with the previous estimation of $Q_\mr{turb}\sim10^{-26}~\mr{erg~cm^{-3}~s^{-1}}$ \citep{zhuravleva14}, although these two calculations are performed on the same target but in a completely different way: the previous estimation by \citet{zhuravleva14} has been done using the surface brightness fluctuations in a statistical manner, while we estimate it thermodynamically from a single distinct substructure. This may indicate the importance of turbulent heating regarding the cooling problem of cluster cores.

The result indicates that the turbulence triggered by sloshing-induced KHIs may have nonnegligible contributions to the ICM turbulence, which has not been considered extensively based on the observational perspective. It is possible that the gravitational energy injected by minor mergers supports the heat input into the ICM because gas sloshing is easily triggered by minor mergers that are constantly happening during the growth of galaxy clusters. We though note that there is no way to regulate heat input from minor mergers to exactly balance cooling, and thus it cannot be the only mechanism.

Recently, \citet{hitomi16} and \citet{hitomiv} have measured the line-of-sight (LOS) velocity dispersion of the gas in the core of the Perseus cluster at $\sim100-200\mr{km~s^{-1}}$. Our estimated turbulent strength is higher than this value. However, there are many uncertainties regarding the comparison; firstly, the region that we used for the turbulent estimation is not covered by the {\it Hitomi} observations. Also, the turbulent scales that the two observations probed may be different from each other. Secondly, our estimated value is based on the assumption of isotropy, which is probably not correct because KHIs occur along a shear flow, which has a certain direction (in the plane of the sky and not in the LOS in our case). Thirdly, how each turbulent component that originates from each turbulent driver (e.g. AGN or sloshing) contributes to the total turbulence may be different between the two observations.

\subsubsection{Convergent flows}
Although the azimuthal average of the temperature and the entropy profiles (see Figure~\ref{img:perseus_prof_eastcf}) increase monotonically toward the larger radii, the profiles along the double edge considered here do not seem to follow them. Instead, the temperature and the entropy profiles seem flat beyond the second break. At the same time, they may also indicate a signature of flattening between the first and the second break (1.4--1.6~arcmin).

These profiles indicate the existence of convergent gas flows at the second break, where the hot and high-entropy gas is moving inward from the outside and the cold and low-entropy gas is moving outward from the inside. Such convergent gas flows around sloshing cold fronts have been observationally suggested \citep[][]{werner15} and also actually indicated in the numerical simulations \citep[e.g.][]{ascasibar06,roediger11}. The reason of the flattening of the thermodynamic profiles inside the second break being not as clear as those outside the second break may be the complex gas flow due to the developing KH eddies. The pressure enhancement above the second break may also be related to the compression due to convergent flows.

\subsubsection{Weak shock}
Inside the first break around 1.2~arcmin, a pressure break is seen. It is also seen independently in temperature, density and entropy profiles. The higher-temperature side has the higher density, which indicates that this is a weak shock front. It is possible that this feature is related to the sloshing cold front in some way, but it is more likely that it is related to the inner ripple-like structures.

\subsection{Brightness dip}

As shown in the optical image in Figure~\ref{img:perseus_unsharp_residual_zoom}, we do not see any corresponding optical structures around the brightness dip region, suggesting the brightness dip is purely an ICM substructure, i.e., not a structure generated by stars or galaxies, but simply due to the distribution of the diffuse gas. The temperatures of the three regions agree with each other (see Table~\ref{tbl:perseus_prof_dip_1t}), which indicates that the brightness dip region is unlikely to be associated with a temperature structure such as shock-heated or adiabatically compressed gas.

The brightness dip is apparently dark, and the fitting results suggest that the normalization at the dip is lower than the surroundings. Therefore, we suggest that the brightness dip is most probably a region where the gas, whose properties are similar to the surroundings, is simply depleted in terms of the line-of-sight volume or the density.

\subsubsection{Magnetic field strength}
Based on the gas depletion scenario, here we investigate the microphysical properties of the ICM. As shown in Figure~\ref{img:perseus_unsharp_residual}, the feather-like structures exist just beneath the western cold front. We point out the similarity of these structures to the recent numerical simulations of the gas sloshing of magnetized plasma \citep[e.g.][]{zuhone11,zuhone15,werner15}. The simulations suggest that when a tangential flow due to the sloshing motion exists and the plasma is magnetized, the magnetic fields therein are stretched and amplified along the flow direction even if the magnetic fields were initially tangled. The stretched magnetic fields push out the gas around them with the amplified magnetic pressure, resulting in a fluctuation of the surface brightness which represents the alignment of the magnetic fields inside the projected volume of the ICM.

Assuming that the ICM is uniform in each region and has a line-of-sight depth of $L$, the deficit of the area-normalized \verb+apec+ normalization of $\sim$0.4$\times$10$^{-6}$ shown in Table~\ref{tbl:perseus_prof_dip_1t} is translated to the deficit of the electron density of $\Delta n_e\sim$0.002~cm$^{-3}(L/42~\mr{kpc})^{-1/2}$, where 42~kpc is the distance of the brightness dip from the cluster centre ($\sim$2~arcmin). Given the temperature uniformity, the physical density deficit directly indicates the deficit of the electron pressure $\Delta p_e \sim 0.01~\mr{keV~cm}^{-3}(L/42~\mr{kpc})^{-1/2}(kT/3.9~\mr{keV})$, where $kT$ is the gas temperature, which should be supported by some other pressure component(s) different from thermal pressure.

Considering the apparent similarity between the Perseus cluster (e.g. Figure~\ref{img:perseus_unsharp_residual}) and the numerical simulation result by \citet{werner15} as well as other results of magnetized gas sloshing simulations in the literature \citep[e.g.][]{zuhone11,zuhone13}, the most natural physical mechanism operating to support the gas pressure is magnetic pressure. The $\Delta p = B^2/8\pi$ relation immediately yields the magnetic field strength $B \sim30~\mu\mr{G}(L/42~\mr{kpc})^{-1/4}(kT/3.9~\mr{keV})^{1/2}$.

The estimated magnetic field strength of $\sim30~\mu\mr{G}$ seems rather high, considering that the ambient magnetic field in the ICM has been estimated at around several $\mu\mr{G}$ \citep{carilli02,bonafede10,bonafede13}. However, there have been simulations which suggest that the ambient magnetic field can be amplified by some factors from several $\mu\mr{G}$, especially in the clusters which show dynamical activity such as gas sloshing \citep[e.g.][]{zuhone11,zuhone13,zuhone15}. Actually this value is measured only for the dip and not representative of average magnetic field strength in bulk of ICM, and given that the Perseus cluster indeed shows many signs of dynamical activity we think this estimation is not implausible. Note that even for the ambient values, there have been observations which indicate ambient ICM magnetic field strengths of $\lesssim40\mu\mr{G}$ \citep[e.g.][]{taylor93,allen01,carilli02}. Also, there have been observations which indicate magnetic field strengths of $20-70\mu\mr{G}$ \citep[e.g.][]{taylor07,fabian08,werner13a} in H$\alpha$ filaments. 

\subsubsection{Other candidates for the pressure support}
The above scenario where the brightness dip is attributed to the magnetic fields amplified due to the gas motion is consistent with previous studies and simulations. However, other sources of pressure, such as non-thermal pressure support by relativistic particles or turbulent pressure support by gas motions, may also play a role. Importantly, we cannot reject some of these other scenarios, at least with the current observations.

Current radio data \citep[70--600~MHz, compiled in ][]{gendron-marsolais17} do not show any clear distinct feature at this location. However, one viable alternative scenario is that the brightness dip is a ghost bubble which represents a past activity of the central active galaxy, similarly to the other brightness cavities shown in Figure~\ref{img:perseus_unsharp_residual}. Since such ghost cavities simply push out the gas, the resulting thermodynamic structure should be similar to the case of magnetic field amplification. However, given its relatively small size compared to the other bubbles, we do not think this is the case, because if the gas depletion were due to the ghost cavity, it would have had time to expand to the size similar to the other cavities, during its buoyant uplift.

Turbulence is another candidate, but it seems unlikely that turbulence is localized to within such a small and clearly confined region. We can in principle test this scenario by measuring the width of the emission lines using high-resolution spectroscopy.

\section{Conclusions}
In this paper, we studied substructures associated with the sloshing cold front in the core of the Perseus cluster using $\sim$1~Msec archival {\it Chandra} ACIS-S data. The main results of our work are summarized below.
\begin{enumerate}
 \item We find that the west half of the cold front seems relatively smooth, while the east half of the front exhibits a more complex, double-layered structure. We point out the similarity between this cold front and recent low-viscosity numerical simulations of KHIs in the context of gas sloshing.
 \item We find that the surface brightness profile across the double-layered front has two brightness edges, which is predicted by numerical simulations of KHIs along the sloshing cold front. We measure the thickness of the layer to be $8.8\pm0.4~\mr{kpc}$ and the azimuthal extension of the layer to be 30--40~kpc, whose ratio is also consistent with the prediction from the numerical simulation.
 \item We find that the thermodynamic structure across the double-layered front is consistent with it being a developing KHI layer along the sloshing cold front.
 \item We find a pressure deficit in the thermodynamic profile at the corresponding radii of the KHI layer candidate. Assuming the line-of-sight geometry, we estimated that the pressure deficit is of the order of $10^{-2}~\mr{keV~cm^{-3}}$.
 \item If the pressure is fully supported by turbulent pressure, the turbulent strength is estimated at $V_{1\mr{d}}\sim400~\mr{km~s^{-1}}$, which is within an order of magnitude of previous estimations using other complementary methods. Assuming that the current size of the structure and the current turbulent strength represent the driving scale of the turbulence, we estimated the turbulent heating rate at $Q_\mr{turb}\sim3\times10^{-26}~\mr{erg~cm^{-3}~s^{-1}}$, which can balance the radiative cooling at this radius. It indicates the importance of turbulent heating regarding the cluster cooling problem, and at the same time that the turbulence triggered by sloshing-induced KHIs may have nonnegligible contribution to the ICM turbulence.
 \item We find feather-like structures underneath the west half of the front, which are similar to the structures that emerge in the recent numerical simulations of the gas sloshing of magnetized plasma.
 \item The thermodynamic properties of the brightness dip, the clearest of the feather-like structures, are consistent with it being the projected gas depletion layer induced by the amplified magnetic field.
 \item Based on this scenario, we estimated the amplified magnetic field strength at $B\sim$30~$\mu\mr{G}$.
\end{enumerate}

\section*{Acknowledgements}
YI is supported by Rikkyo University Special Fund for Research (SFR). AS gratefully acknowledges support by the Women In Science Excel (WISE) programme of the Netherlands Organisation for Scientific Research (NWO), and is thankful to the Kavli Institute for the Physics and Mathematics of the Universe for their continued hospitality. NW is supported by the Lend{\"u}let LP2016-11 grant awarded by the Hungarian Academy of Sciences. ACF acknowledges support from ERC Advanced Grant 340442.




\begin{thebibliography}{}
\makeatletter
\relax
\def\mn@urlcharsother{\let\do\@makeother \do\$\do\&\do\#\do\^\do\_\do\%\do\~}
\def\mn@doi{\begingroup\mn@urlcharsother \@ifnextchar [ {\mn@doi@}
  {\mn@doi@[]}}
\def\mn@doi@[#1]#2{\def\@tempa{#1}\ifx\@tempa\@empty \href
  {http://dx.doi.org/#2} {doi:#2}\else \href {http://dx.doi.org/#2} {#1}\fi
  \endgroup}
\def\mn@eprint#1#2{\mn@eprint@#1:#2::\@nil}
\def\mn@eprint@arXiv#1{\href {http://arxiv.org/abs/#1} {{\tt arXiv:#1}}}
\def\mn@eprint@dblp#1{\href {http://dblp.uni-trier.de/rec/bibtex/#1.xml}
  {dblp:#1}}
\def\mn@eprint@#1:#2:#3:#4\@nil{\def\@tempa {#1}\def\@tempb {#2}\def\@tempc
  {#3}\ifx \@tempc \@empty \let \@tempc \@tempb \let \@tempb \@tempa \fi \ifx
  \@tempb \@empty \def\@tempb {arXiv}\fi \@ifundefined
  {mn@eprint@\@tempb}{\@tempb:\@tempc}{\expandafter \expandafter \csname
  mn@eprint@\@tempb\endcsname \expandafter{\@tempc}}}

\bibitem[\protect\citeauthoryear{{Ahn} et~al.,}{{Ahn} et~al.}{2014}]{ahn14}
{Ahn} C.~P.,  et~al., 2014, \mn@doi [\apjs] {10.1088/0067-0049/211/2/17}, \href
  {http://ads.nao.ac.jp/abs/2014ApJS..211...17A} {211, 17}

\bibitem[\protect\citeauthoryear{{Allen} et~al.,}{{Allen}
  et~al.}{2001}]{allen01}
{Allen} S.~W.,  et~al., 2001, \mn@doi [\mnras]
  {10.1046/j.1365-8711.2001.04315.x}, \href
  {http://ads.nao.ac.jp/abs/2001MNRAS.324..842A} {324, 842}

\bibitem[\protect\citeauthoryear{{Ascasibar} \& {Markevitch}}{{Ascasibar} \&
  {Markevitch}}{2006}]{ascasibar06}
{Ascasibar} Y.,  {Markevitch} M.,  2006, \mn@doi [\apj] {10.1086/506508}, \href
  {http://ads.nao.ac.jp/abs/2006ApJ...650..102A} {650, 102}

\bibitem[\protect\citeauthoryear{{Barcons}, {Nandra}, {Barret}, {den Herder},
  {Fabian}, {Piro}, {Watson}  \& {the Athena Team}}{{Barcons}
  et~al.}{2015}]{barcons15}
{Barcons} X.,  {Nandra} K.,  {Barret} D.,  {den Herder} J.-W.,  {Fabian} A.~C.,
   {Piro} L.,  {Watson} M.~G.,   {the Athena Team} 2015, \mn@doi [Journal of
  Physics Conference Series] {10.1088/1742-6596/610/1/012008}, \href
  {http://adsabs.harvard.edu/abs/2015JPhCS.610a2008B} {610, 012008}

\bibitem[\protect\citeauthoryear{{Boehringer}, {Voges}, {Fabian}, {Edge}  \&
  {Neumann}}{{Boehringer} et~al.}{1993}]{boehringer93}
{Boehringer} H.,  {Voges} W.,  {Fabian} A.~C.,  {Edge} A.~C.,   {Neumann}
  D.~M.,  1993, \mnras, \href {http://ads.nao.ac.jp/abs/1993MNRAS.264L..25B}
  {264, L25}

\bibitem[\protect\citeauthoryear{{Bonafede}, {Feretti}, {Murgia}, {Govoni},
  {Giovannini}, {Dallacasa}, {Dolag}  \& {Taylor}}{{Bonafede}
  et~al.}{2010}]{bonafede10}
{Bonafede} A.,  {Feretti} L.,  {Murgia} M.,  {Govoni} F.,  {Giovannini} G.,
  {Dallacasa} D.,  {Dolag} K.,   {Taylor} G.~B.,  2010, \mn@doi [\aap]
  {10.1051/0004-6361/200913696}, \href
  {http://ads.nao.ac.jp/abs/2010A%26A...513A..30B} {513, A30}

\bibitem[\protect\citeauthoryear{{Bonafede}, {Vazza}, {Br{\"u}ggen}, {Murgia},
  {Govoni}, {Feretti}, {Giovannini}  \& {Ogrean}}{{Bonafede}
  et~al.}{2013}]{bonafede13}
{Bonafede} A.,  {Vazza} F.,  {Br{\"u}ggen} M.,  {Murgia} M.,  {Govoni} F.,
  {Feretti} L.,  {Giovannini} G.,   {Ogrean} G.,  2013, \mn@doi [\mnras]
  {10.1093/mnras/stt960}, \href {http://ads.nao.ac.jp/abs/2013MNRAS.433.3208B}
  {433, 3208}

\bibitem[\protect\citeauthoryear{{Carilli} \& {Taylor}}{{Carilli} \&
  {Taylor}}{2002}]{carilli02}
{Carilli} C.~L.,  {Taylor} G.~B.,  2002, \mn@doi [\araa]
  {10.1146/annurev.astro.40.060401.093852}, \href
  {http://ads.nao.ac.jp/abs/2002ARA%26A..40..319C} {40, 319}

\bibitem[\protect\citeauthoryear{{Churazov}, {Forman}, {Jones}  \&
  {B{\"o}hringer}}{{Churazov} et~al.}{2000}]{churazov00}
{Churazov} E.,  {Forman} W.,  {Jones} C.,   {B{\"o}hringer} H.,  2000, \aap,
  \href {http://ads.nao.ac.jp/abs/2000A%26A...356..788C} {356, 788}

\bibitem[\protect\citeauthoryear{{Churazov}, {Forman}, {Jones}  \&
  {B{\"o}hringer}}{{Churazov} et~al.}{2003}]{churazov03}
{Churazov} E.,  {Forman} W.,  {Jones} C.,   {B{\"o}hringer} H.,  2003, \mn@doi
  [\apj] {10.1086/374923}, \href {http://ads.nao.ac.jp/abs/2003ApJ...590..225C}
  {590, 225}

\bibitem[\protect\citeauthoryear{{Edge}, {Stewart}, {Fabian}  \&
  {Arnaud}}{{Edge} et~al.}{1990}]{edge90}
{Edge} A.~C.,  {Stewart} G.~C.,  {Fabian} A.~C.,   {Arnaud} K.~A.,  1990,
  \mnras, \href {http://ads.nao.ac.jp/abs/1990MNRAS.245..559E} {245, 559}

\bibitem[\protect\citeauthoryear{{Eisenstein} et~al.,}{{Eisenstein}
  et~al.}{2011}]{eisenstein11}
{Eisenstein} D.~J.,  et~al., 2011, \mn@doi [\aj] {10.1088/0004-6256/142/3/72},
  \href {http://adsabs.harvard.edu/abs/2011AJ....142...72E} {142, 72}

\bibitem[\protect\citeauthoryear{{Fabian} et~al.,}{{Fabian}
  et~al.}{2000}]{fabian00}
{Fabian} A.~C.,  et~al., 2000, \mn@doi [\mnras]
  {10.1046/j.1365-8711.2000.03904.x}, \href
  {http://ads.nao.ac.jp/abs/2000MNRAS.318L..65F} {318, L65}

\bibitem[\protect\citeauthoryear{{Fabian}, {Sanders}, {Taylor}, {Allen},
  {Crawford}, {Johnstone}  \& {Iwasawa}}{{Fabian} et~al.}{2006}]{fabian06}
{Fabian} A.~C.,  {Sanders} J.~S.,  {Taylor} G.~B.,  {Allen} S.~W.,  {Crawford}
  C.~S.,  {Johnstone} R.~M.,   {Iwasawa} K.,  2006, \mn@doi [\mnras]
  {10.1111/j.1365-2966.2005.09896.x}, \href
  {http://ads.nao.ac.jp/abs/2006MNRAS.366..417F} {366, 417}

\bibitem[\protect\citeauthoryear{{Fabian}, {Johnstone}, {Sanders}, {Conselice},
  {Crawford}, {Gallagher}  \& {Zweibel}}{{Fabian} et~al.}{2008}]{fabian08}
{Fabian} A.~C.,  {Johnstone} R.~M.,  {Sanders} J.~S.,  {Conselice} C.~J.,
  {Crawford} C.~S.,  {Gallagher} III J.~S.,   {Zweibel} E.,  2008, \mn@doi
  [\nat] {10.1038/nature07169}, \href
  {http://ads.nao.ac.jp/abs/2008Natur.454..968F} {454, 968}

\bibitem[\protect\citeauthoryear{{Fabian} et~al.,}{{Fabian}
  et~al.}{2011}]{fabian11b}
{Fabian} A.~C.,  et~al., 2011, \mn@doi [\mnras]
  {10.1111/j.1365-2966.2011.19402.x}, \href
  {http://ads.nao.ac.jp/abs/2011MNRAS.418.2154F} {418, 2154}

\bibitem[\protect\citeauthoryear{{Gendron-Marsolais}
  et~al.,}{{Gendron-Marsolais} et~al.}{2017}]{gendron-marsolais17}
{Gendron-Marsolais} M.,  et~al., 2017, \mn@doi [\mnras]
  {10.1093/mnras/stx1042}, \href
  {http://adsabs.harvard.edu/abs/2017MNRAS.469.3872G} {469, 3872}

\bibitem[\protect\citeauthoryear{{Hitomi Collaboration} et~al.,}{{Hitomi
  Collaboration} et~al.}{2016}]{hitomi16}
{Hitomi Collaboration} et~al., 2016, \mn@doi [\nat] {10.1038/nature18627},
  \href {http://ads.nao.ac.jp/abs/2016Natur.535..117H} {535, 117}

\bibitem[\protect\citeauthoryear{{Hitomi Collaboration} et~al.,}{{Hitomi
  Collaboration} et~al.}{2018}]{hitomiv}
{Hitomi Collaboration} et~al., 2018, \mn@doi [\pasj] {10.1093/pasj/psx138},
  \href {http://adsabs.harvard.edu/abs/2018PASJ...70....9H} {70, 9}

\bibitem[\protect\citeauthoryear{Ichinohe, Werner, Simionescu, Allen, Canning,
  Ehlert, Mernier  \& Takahashi}{Ichinohe et~al.}{2015}]{ichinohe15}
Ichinohe Y.,  Werner N.,  Simionescu A.,  Allen S.~W.,  Canning R. E.~A.,
  Ehlert S.,  Mernier F.,   Takahashi T.,  2015, Monthly Notices of the Royal
  Astronomical Society, 448, 2971

\bibitem[\protect\citeauthoryear{{Ichinohe}, {Simionescu}, {Werner}  \&
  {Takahashi}}{{Ichinohe} et~al.}{2017}]{ichinohe17}
{Ichinohe} Y.,  {Simionescu} A.,  {Werner} N.,   {Takahashi} T.,  2017, \mn@doi
  [\mnras] {10.1093/mnras/stx280}, \href
  {http://adsabs.harvard.edu/abs/2017MNRAS.467.3662I} {467, 3662}

\bibitem[\protect\citeauthoryear{{Kalberla}, {Burton}, {Hartmann}, {Arnal},
  {Bajaja}, {Morras}  \& {P{\"o}ppel}}{{Kalberla} et~al.}{2005}]{kalberla05}
{Kalberla} P.~M.~W.,  {Burton} W.~B.,  {Hartmann} D.,  {Arnal} E.~M.,  {Bajaja}
  E.,  {Morras} R.,   {P{\"o}ppel} W.~G.~L.,  2005, \mn@doi [\aap]
  {10.1051/0004-6361:20041864}, \href
  {http://ads.nao.ac.jp/abs/2005A%26A...440..775K} {440, 775}

\bibitem[\protect\citeauthoryear{{Kelley} et~al.,}{{Kelley}
  et~al.}{2016}]{kelley16}
{Kelley} R.~L.,  et~al., 2016, in Society of Photo-Optical Instrumentation
  Engineers (SPIE) Conference Series. p. 99050V, \mn@doi{10.1117/12.2232509}

\bibitem[\protect\citeauthoryear{{Keshet}, {Markevitch}, {Birnboim}  \&
  {Loeb}}{{Keshet} et~al.}{2010}]{keshet10}
{Keshet} U.,  {Markevitch} M.,  {Birnboim} Y.,   {Loeb} A.,  2010, \mn@doi
  [\apjl] {10.1088/2041-8205/719/1/L74}, \href
  {http://adsabs.harvard.edu/abs/2010ApJ...719L..74K} {719, L74}

\bibitem[\protect\citeauthoryear{{Markevitch} \& {Vikhlinin}}{{Markevitch} \&
  {Vikhlinin}}{2007}]{markevitch07}
{Markevitch} M.,  {Vikhlinin} A.,  2007, \mn@doi [\physrep]
  {10.1016/j.physrep.2007.01.001}, \href
  {http://ads.nao.ac.jp/abs/2007PhR...443....1M} {443, 1}

\bibitem[\protect\citeauthoryear{{Reiss} \& {Keshet}}{{Reiss} \&
  {Keshet}}{2014}]{reiss14}
{Reiss} I.,  {Keshet} U.,  2014, \mn@doi [Physical Review Letters]
  {10.1103/PhysRevLett.113.071302}, \href
  {http://adsabs.harvard.edu/abs/2014PhRvL.113g1302R} {113, 071302}

\bibitem[\protect\citeauthoryear{{Roediger}, {Br{\"u}ggen}, {Simionescu},
  {B{\"o}hringer}, {Churazov}  \& {Forman}}{{Roediger}
  et~al.}{2011}]{roediger11}
{Roediger} E.,  {Br{\"u}ggen} M.,  {Simionescu} A.,  {B{\"o}hringer} H.,
  {Churazov} E.,   {Forman} W.~R.,  2011, \mn@doi [\mnras]
  {10.1111/j.1365-2966.2011.18279.x}, \href
  {http://ads.nao.ac.jp/abs/2011MNRAS.413.2057R} {413, 2057}

\bibitem[\protect\citeauthoryear{{Roediger}, {Kraft}, {Nulsen}, {Churazov},
  {Forman}, {Br{\"u}ggen}  \& {Kokotanekova}}{{Roediger}
  et~al.}{2013a}]{roediger13b}
{Roediger} E.,  {Kraft} R.~P.,  {Nulsen} P.,  {Churazov} E.,  {Forman} W.,
  {Br{\"u}ggen} M.,   {Kokotanekova} R.,  2013a, \mn@doi [\mnras]
  {10.1093/mnras/stt1691}, \href {http://ads.nao.ac.jp/abs/2013MNRAS.436.1721R}
  {436, 1721}

\bibitem[\protect\citeauthoryear{{Roediger}, {Kraft}, {Forman}, {Nulsen}  \&
  {Churazov}}{{Roediger} et~al.}{2013b}]{roediger13a}
{Roediger} E.,  {Kraft} R.~P.,  {Forman} W.~R.,  {Nulsen} P.~E.~J.,
  {Churazov} E.,  2013b, \mn@doi [\apj] {10.1088/0004-637X/764/1/60}, \href
  {http://ads.nao.ac.jp/abs/2013ApJ...764...60R} {764, 60}

\bibitem[\protect\citeauthoryear{{Sanders}}{{Sanders}}{2006}]{sanders06}
{Sanders} J.~S.,  2006, \mn@doi [\mnras] {10.1111/j.1365-2966.2006.10716.x},
  \href {http://ads.nao.ac.jp/abs/2006MNRAS.371..829S} {371, 829}

\bibitem[\protect\citeauthoryear{{Sanders} \& {Fabian}}{{Sanders} \&
  {Fabian}}{2007}]{sanders07}
{Sanders} J.~S.,  {Fabian} A.~C.,  2007, \mn@doi [\mnras]
  {10.1111/j.1365-2966.2007.12347.x}, \href
  {http://ads.nao.ac.jp/abs/2007MNRAS.381.1381S} {381, 1381}

\bibitem[\protect\citeauthoryear{{Sanders}, {Fabian}, {Russell}, {Walker}  \&
  {Blundell}}{{Sanders} et~al.}{2016}]{sanders16b}
{Sanders} J.~S.,  {Fabian} A.~C.,  {Russell} H.~R.,  {Walker} S.~A.,
  {Blundell} K.~M.,  2016, \mn@doi [\mnras] {10.1093/mnras/stw1119}, \href
  {http://adsabs.harvard.edu/abs/2016MNRAS.460.1898S} {460, 1898}

\bibitem[\protect\citeauthoryear{{Simionescu} et~al.,}{{Simionescu}
  et~al.}{2011}]{simionescu11}
{Simionescu} A.,  et~al., 2011, \mn@doi [Science] {10.1126/science.1200331},
  \href {http://ads.nao.ac.jp/abs/2011Sci...331.1576S} {331, 1576}

\bibitem[\protect\citeauthoryear{{Simionescu} et~al.,}{{Simionescu}
  et~al.}{2012}]{simionescu12}
{Simionescu} A.,  et~al., 2012, \mn@doi [\apj] {10.1088/0004-637X/757/2/182},
  \href {http://ads.nao.ac.jp/abs/2012ApJ...757..182S} {757, 182}

\bibitem[\protect\citeauthoryear{{Su} et~al.,}{{Su} et~al.}{2017}]{su17}
{Su} Y.,  et~al., 2017, \mn@doi [\apj] {10.3847/1538-4357/834/1/74}, \href
  {http://adsabs.harvard.edu/abs/2017ApJ...834...74S} {834, 74}

\bibitem[\protect\citeauthoryear{{Sutherland} \& {Dopita}}{{Sutherland} \&
  {Dopita}}{1993}]{sutherland93}
{Sutherland} R.~S.,  {Dopita} M.~A.,  1993, \mn@doi [\apjs] {10.1086/191823},
  \href {http://ads.nao.ac.jp/abs/1993ApJS...88..253S} {88, 253}

\bibitem[\protect\citeauthoryear{{Takahashi}, {Kokubun}, {Mitsuda}  \& et
  al.}{{Takahashi} et~al.}{2016}]{takahashi16}
{Takahashi} T.,  {Kokubun} M.,  {Mitsuda} K.,   et al. 2016, in Society of
  Photo-Optical Instrumentation Engineers (SPIE) Conference Series. p. 99050U,
  \mn@doi{10.1117/12.2232379}

\bibitem[\protect\citeauthoryear{{Tashiro} et~al.,}{{Tashiro}
  et~al.}{2018}]{tashiro18}
{Tashiro} M.,  et~al., 2018, in Society of Photo-Optical Instrumentation
  Engineers (SPIE) Conference Series. p. 1069922, \mn@doi{10.1117/12.2309455}

\bibitem[\protect\citeauthoryear{{Taylor} \& {Perley}}{{Taylor} \&
  {Perley}}{1993}]{taylor93}
{Taylor} G.~B.,  {Perley} R.~A.,  1993, \mn@doi [\apj] {10.1086/173257}, \href
  {http://ads.nao.ac.jp/abs/1993ApJ...416..554T} {416, 554}

\bibitem[\protect\citeauthoryear{{Taylor}, {Fabian}, {Gentile}, {Allen},
  {Crawford}  \& {Sanders}}{{Taylor} et~al.}{2007}]{taylor07}
{Taylor} G.~B.,  {Fabian} A.~C.,  {Gentile} G.,  {Allen} S.~W.,  {Crawford} C.,
    {Sanders} J.~S.,  2007, \mn@doi [\mnras]
  {10.1111/j.1365-2966.2007.12368.x}, \href
  {http://ads.nao.ac.jp/abs/2007MNRAS.382...67T} {382, 67}

\bibitem[\protect\citeauthoryear{{Ueda}, {Kitayama}  \& {Dotani}}{{Ueda}
  et~al.}{2017}]{ueda17}
{Ueda} S.,  {Kitayama} T.,   {Dotani} T.,  2017, \mn@doi [\apj]
  {10.3847/1538-4357/aa5c3e}, \href
  {http://adsabs.harvard.edu/abs/2017ApJ...837...34U} {837, 34}

\bibitem[\protect\citeauthoryear{{Urban} et~al.,}{{Urban}
  et~al.}{2014}]{urban14}
{Urban} O.,  et~al., 2014, \mn@doi [\mnras] {10.1093/mnras/stt2209}, \href
  {http://ads.nao.ac.jp/abs/2014MNRAS.437.3939U} {437, 3939}

\bibitem[\protect\citeauthoryear{{Walker}, {Hlavacek-Larrondo},
  {Gendron-Marsolais}, {Fabian}, {Intema}, {Sanders}, {Bamford}  \& {van
  Weeren}}{{Walker} et~al.}{2017}]{walker17}
{Walker} S.~A.,  {Hlavacek-Larrondo} J.,  {Gendron-Marsolais} M.,  {Fabian}
  A.~C.,  {Intema} H.,  {Sanders} J.~S.,  {Bamford} J.~T.,   {van Weeren} R.,
  2017, \mn@doi [\mnras] {10.1093/mnras/stx640}, \href
  {http://adsabs.harvard.edu/abs/2017MNRAS.468.2506W} {468, 2506}

\bibitem[\protect\citeauthoryear{{Walker}, {ZuHone}, {Fabian}  \&
  {Sanders}}{{Walker} et~al.}{2018}]{walker18}
{Walker} S.~A.,  {ZuHone} J.,  {Fabian} A.,   {Sanders} J.,  2018, \mn@doi
  [Nature Astronomy] {10.1038/s41550-018-0401-8}, \href
  {http://adsabs.harvard.edu/abs/2018NatAs...2..292W} {2, 292}

\bibitem[\protect\citeauthoryear{{Wang} \& {Markevitch}}{{Wang} \&
  {Markevitch}}{2018}]{wang18}
{Wang} Q.,  {Markevitch} M.,  2018, preprint, \href
  {http://adsabs.harvard.edu/abs/2018arXiv181002813W} {} (\mn@eprint {arXiv}
  {1810.02813})

\bibitem[\protect\citeauthoryear{{Werner} et~al.,}{{Werner}
  et~al.}{2013}]{werner13a}
{Werner} N.,  et~al., 2013, \mn@doi [\apj] {10.1088/0004-637X/767/2/153}, \href
  {http://ads.nao.ac.jp/abs/2013ApJ...767..153W} {767, 153}

\bibitem[\protect\citeauthoryear{{Werner} et~al.,}{{Werner}
  et~al.}{2016a}]{werner15}
{Werner} N.,  et~al., 2016a, \mn@doi [\mnras] {10.1093/mnras/stv2358}, \href
  {http://ads.nao.ac.jp/abs/2016MNRAS.455..846W} {455, 846}

\bibitem[\protect\citeauthoryear{{Werner} et~al.,}{{Werner}
  et~al.}{2016b}]{werner16}
{Werner} N.,  et~al., 2016b, \mn@doi [\mnras] {10.1093/mnras/stw1171}, \href
  {http://ads.nao.ac.jp/abs/2016MNRAS.460.2752W} {460, 2752}

\bibitem[\protect\citeauthoryear{{Zhuravleva} et~al.,}{{Zhuravleva}
  et~al.}{2014}]{zhuravleva14}
{Zhuravleva} I.,  et~al., 2014, \mn@doi [\nat] {10.1038/nature13830}, \href
  {http://ads.nao.ac.jp/abs/2014Natur.515...85Z} {515, 85}

\bibitem[\protect\citeauthoryear{{ZuHone}, {Markevitch}  \& {Lee}}{{ZuHone}
  et~al.}{2011}]{zuhone11}
{ZuHone} J.~A.,  {Markevitch} M.,   {Lee} D.,  2011, \mn@doi [\apj]
  {10.1088/0004-637X/743/1/16}, \href
  {http://ads.nao.ac.jp/abs/2011ApJ...743...16Z} {743, 16}

\bibitem[\protect\citeauthoryear{{ZuHone}, {Markevitch}, {Ruszkowski}  \&
  {Lee}}{{ZuHone} et~al.}{2013}]{zuhone13}
{ZuHone} J.~A.,  {Markevitch} M.,  {Ruszkowski} M.,   {Lee} D.,  2013, \mn@doi
  [\apj] {10.1088/0004-637X/762/2/69}, \href
  {http://ads.nao.ac.jp/abs/2013ApJ...762...69Z} {762, 69}

\bibitem[\protect\citeauthoryear{{ZuHone}, {Kunz}, {Markevitch}, {Stone}  \&
  {Biffi}}{{ZuHone} et~al.}{2015}]{zuhone15}
{ZuHone} J.~A.,  {Kunz} M.~W.,  {Markevitch} M.,  {Stone} J.~M.,   {Biffi} V.,
  2015, \mn@doi [\apj] {10.1088/0004-637X/798/2/90}, \href
  {http://ads.nao.ac.jp/abs/2015ApJ...798...90Z} {798, 90}

\bibitem[\protect\citeauthoryear{{Zuhone} \& {Roediger}}{{Zuhone} \&
  {Roediger}}{2016}]{zuhone16}
{Zuhone} J.~A.,  {Roediger} E.,  2016, \mn@doi [Journal of Plasma Physics]
  {10.1017/S0022377816000544}, \href
  {https://ui.adsabs.harvard.edu/#abs/2016JPlPh..82c5301Z} {82, 535820301}

\makeatother
\end{thebibliography}






\bsp	
\label{lastpage}
\end{document}